\documentclass[preprint,amsmath,amssymb,aps,prd,superscriptaddress,floatfix,nofootinbib]{revtex4-2}

\usepackage[top=2.5cm,bottom=1.8cm,left=2.6cm,right=2.5cm]{geometry}
\usepackage[utf8]{inputenc}
\usepackage[T1]{fontenc}
\usepackage{subcaption}
\usepackage{graphicx}
\usepackage{bm}

\usepackage[final]{changes} 
\colorlet{Changes@Color}{red}

\DeclareMathAlphabet{\mathsfit}{\encodingdefault}{\sfdefault}{m}{sl}
\SetMathAlphabet{\mathsfit}{bold}{\encodingdefault}{\sfdefault}{bx}{n}
\newcommand{\tens}[1]{\bm{\mathsfit{#1}}}
\def\tX{{\tens{X}}}

\begin{document}

\title{Neutrino interaction classification with a convolutional neural network in the DUNE far detector}

\begin{abstract}{The Deep Underground Neutrino Experiment is a next-generation neutrino oscillation experiment that aims to measure $CP$-violation in the neutrino sector as part of a wider physics program. A deep learning approach based on a convolutional neural network has been developed to provide highly efficient and pure selections of electron neutrino and muon neutrino charged-current interactions. The electron neutrino (antineutrino) selection efficiency peaks at 90\% (94\%) and exceeds 85\% (90\%) for reconstructed neutrino energies between 2-5\,GeV. The muon neutrino (antineutrino) event selection is found to have a maximum efficiency of 96\% (97\%) and exceeds 90\% (95\%) efficiency for reconstructed neutrino energies above 2\,GeV. When considering all electron neutrino and antineutrino interactions as signal, a selection purity of 90\% is achieved. These event selections are critical to maximize the sensitivity of the experiment to $CP$-violating effects.}
\end{abstract}


\newcommand{\Amsterdam}{University of Amsterdam, NL-1098 XG Amsterdam, The Netherlands}
\newcommand{\Antananarivo}{University of Antananarivo, Antananarivo 101, Madagascar}
\newcommand{\AntonioNarino}{Universidad Antonio Nari{\~n}o, Bogot{\'a}, Colombia}
\newcommand{\Argonne}{Argonne National Laboratory, Argonne, IL 60439, USA}
\newcommand{\Arizona}{University of Arizona, Tucson, AZ 85721, USA}
\newcommand{\Asuncion}{Universidad Nacional de Asunci{\'o}n, San Lorenzo, Paraguay}
\newcommand{\Athens}{University of Athens, Zografou GR 157 84, Greece}
\newcommand{\Atlantico}{Universidad del Atl{\'a}ntico, Atl{\'a}ntico, Colombia}
\newcommand{\Banaras}{Banaras Hindu University, Varanasi - 221 005, India}
\newcommand{\Basel}{University of Basel, CH-4056 Basel, Switzerland}
\newcommand{\Bern}{University of Bern, CH-3012 Bern, Switzerland}
\newcommand{\Beykent}{Beykent University, Istanbul, Turkey}
\newcommand{\Birmingham}{University of Birmingham, Birmingham B15 2TT, United Kingdom}
\newcommand{\BolognaUniversity}{Universit{\`a} del Bologna, 40127 Bologna, Italy}
\newcommand{\Boston}{Boston University, Boston, MA 02215, USA}
\newcommand{\Bristol}{University of Bristol, Bristol BS8 1TL, United Kingdom}
\newcommand{\Brookhaven}{Brookhaven National Laboratory, Upton, NY 11973, USA}
\newcommand{\Bucharest}{University of Bucharest, Bucharest, Romania}
\newcommand{\CBPF}{Centro Brasileiro de Pesquisas F\'isicas, Rio de Janeiro, RJ 22290-180, Brazil}
\newcommand{\CEASaclay}{CEA/Saclay, IRFU Institut de Recherche sur les Lois Fondamentales de l'Univers, F-91191 Gif-sur-Yvette CEDEX, France}
\newcommand{\CERN}{CERN, The European Organization for Nuclear Research, 1211 Meyrin, Switzerland}
\newcommand{\CIEMAT}{CIEMAT, Centro de Investigaciones Energ{\'e}ticas, Medioambientales y Tecnol{\'o}gicas, E-28040 Madrid, Spain}
\newcommand{\CUSB}{Central University of South Bihar, Gaya {\textendash} 824236, India }
\newcommand{\CalBerkeley}{University of California Berkeley, Berkeley, CA 94720, USA}
\newcommand{\CalDavis}{University of California Davis, Davis, CA 95616, USA}
\newcommand{\CalIrvine}{University of California Irvine, Irvine, CA 92697, USA}
\newcommand{\CalLosangeles}{University of California Los Angeles, Los Angeles, CA 90095, USA}
\newcommand{\CalRiverside}{University of California Riverside, Riverside CA 92521, USA}
\newcommand{\CalSantabarbara}{University of California Santa Barbara, Santa Barbara, California 93106 USA}
\newcommand{\Caltech}{California Institute of Technology, Pasadena, CA 91125, USA}
\newcommand{\Cambridge}{University of Cambridge, Cambridge CB3 0HE, United Kingdom}
\newcommand{\Campinas}{Universidade Estadual de Campinas, Campinas - SP, 13083-970, Brazil}
\newcommand{\CataniaUniversitadi}{Universit{\`a} di Catania, 2 - 95131 Catania, Italy}
\newcommand{\Charles}{Institute of Particle and Nuclear Physics of the Faculty of Mathematics and Physics of the Charles University, 180 00 Prague 8, Czech Republic }
\newcommand{\Chicago}{University of Chicago, Chicago, IL 60637, USA}
\newcommand{\ChungAng}{Chung-Ang University, Seoul 06974, South Korea}
\newcommand{\Cincinnati}{University of Cincinnati, Cincinnati, OH 45221, USA}
\newcommand{\Cinvestav}{Centro de Investigaci{\'o}n y de Estudios Avanzados del Instituto Polit{\'e}cnico Nacional (Cinvestav), Mexico City, Mexico}
\newcommand{\Colima}{Universidad de Colima, Colima, Mexico}
\newcommand{\ColoradoBoulder}{University of Colorado Boulder, Boulder, CO 80309, USA}
\newcommand{\ColoradoState}{Colorado State University, Fort Collins, CO 80523, USA}
\newcommand{\Columbia}{Columbia University, New York, NY 10027, USA}
\newcommand{\CzechAcademyofSciences}{Institute of Physics, Czech Academy of Sciences, 182 00 Prague 8, Czech Republic}
\newcommand{\CzechTechnical}{Czech Technical University, 115 19 Prague 1, Czech Republic}
\newcommand{\DakotaState}{Dakota State University, Madison, SD 57042, USA}
\newcommand{\Dallas}{University of Dallas, Irving, TX 75062-4736, USA}
\newcommand{\DannecyleVieux}{Laboratoire d'Annecy-le-Vieux de Physique des Particules, CNRS/IN2P3 and Universit{\'e} Savoie Mont Blanc, 74941 Annecy-le-Vieux, France}
\newcommand{\Daresbury}{Daresbury Laboratory, Cheshire WA4 4AD, United Kingdom}
\newcommand{\Drexel}{Drexel University, Philadelphia, PA 19104, USA}
\newcommand{\Duke}{Duke University, Durham, NC 27708, USA}
\newcommand{\Durham}{Durham University, Durham DH1 3LE, United Kingdom}
\newcommand{\EIA}{Universidad EIA, Antioquia, Colombia}
\newcommand{\ETH}{ETH Zurich, Zurich, Switzerland}
\newcommand{\Edinburgh}{University of Edinburgh, Edinburgh EH8 9YL, United Kingdom}
\newcommand{\FCULport}{Faculdade de Ci{\^e}ncias da Universidade de Lisboa, Universidade de Lisboa, Portugal}
\newcommand{\FederaldeAlfenas}{Universidade Federal de Alfenas, Po{\c{c}}os de Caldas - MG, 37715-400, Brazil}
\newcommand{\FederaldeGoias}{Universidade Federal de Goias, Goiania, GO 74690-900, Brazil}
\newcommand{\FederaldeSaoCarlos}{Universidade Federal de S{\~a}o Carlos, Araras - SP, 13604-900, Brazil}
\newcommand{\FederaldoABC}{Universidade Federal do ABC, Santo Andr{\'e} - SP, 09210-580 Brazil}
\newcommand{\FederaldoRio}{Universidade Federal do Rio de Janeiro,  Rio de Janeiro - RJ, 21941-901, Brazil}
\newcommand{\Fermi}{Fermi National Accelerator Laboratory, Batavia, IL 60510, USA}
\newcommand{\Florida}{University of Florida, Gainesville, FL 32611-8440, USA}
\newcommand{\Fluminense}{Fluminense Federal University, 9 Icara{\'\i} Niter{\'o}i - RJ, 24220-900, Brazil }
\newcommand{\Genova}{Universit{\`a} degli Studi di Genova, Genova, Italy}
\newcommand{\Georgian}{Georgian Technical University, Tbilisi, Georgia}
\newcommand{\GranSasso}{Gran Sasso Science Institute, L'Aquila, Italy}
\newcommand{\GranSassoLab}{Laboratori Nazionali del Gran Sasso, L'Aquila AQ, Italy}
\newcommand{\Granada}{University of Granada {\&} CAFPE, 18002 Granada, Spain}
\newcommand{\Grenoble}{University Grenoble Alpes, CNRS, Grenoble INP, LPSC-IN2P3, 38000 Grenoble, France}
\newcommand{\Guanajuato}{Universidad de Guanajuato, Guanajuato, C.P. 37000, Mexico}
\newcommand{\Harish}{Harish-Chandra Research Institute, Jhunsi, Allahabad 211 019, India}
\newcommand{\Harvard}{Harvard University, Cambridge, MA 02138, USA}
\newcommand{\Hawaii}{University of Hawaii, Honolulu, HI 96822, USA}
\newcommand{\Houston}{University of Houston, Houston, TX 77204, USA}
\newcommand{\Hyderabad}{University of  Hyderabad, Gachibowli, Hyderabad - 500 046, India}
\newcommand{\IFAE}{Institut de F{\`\i}sica d'Altes Energies, Barcelona, Spain}
\newcommand{\IFIC}{Instituto de Fisica Corpuscular, 46980 Paterna, Valencia, Spain}
\newcommand{\INFNBologna}{Istituto Nazionale di Fisica Nucleare Sezione di Bologna, 40127 Bologna BO, Italy}
\newcommand{\INFNCatania}{Istituto Nazionale di Fisica Nucleare Sezione di Catania, I-95123 Catania, Italy}
\newcommand{\INFNGenova}{Istituto Nazionale di Fisica Nucleare Sezione di Genova, 16146 Genova GE, Italy}
\newcommand{\INFNLecce}{Istituto Nazionale di Fisica Nucleare Sezione di Lecce, 73100 - Lecce, Italy}
\newcommand{\INFNMilanBicocca}{Istituto Nazionale di Fisica Nucleare Sezione di Milano Bicocca, 3 - I-20126 Milano, Italy}
\newcommand{\INFNMilano}{Istituto Nazionale di Fisica Nucleare Sezione di Milano, 20133 Milano, Italy}
\newcommand{\INFNNapoli}{Istituto Nazionale di Fisica Nucleare Sezione di Napoli, I-80126 Napoli, Italy}
\newcommand{\INFNPadova}{Istituto Nazionale di Fisica Nucleare Sezione di Padova, 35131 Padova, Italy}
\newcommand{\INFNPavia}{Istituto Nazionale di Fisica Nucleare Sezione di Pavia,  I-27100 Pavia, Italy}
\newcommand{\INFNSud}{Istituto Nazionale di Fisica Nucleare Laboratori Nazionali del Sud, 95123 Catania, Italy}
\newcommand{\INR}{Institute for Nuclear Research of the Russian Academy of Sciences, Moscow 117312, Russia}
\newcommand{\IPLyon}{Institut de Physique des 2 Infinis de Lyon, 69622 Villeurbanne, France}
\newcommand{\IPM}{Institute for Research in Fundamental Sciences, Tehran, Iran}
\newcommand{\ISTlisboa}{Instituto Superior T{\'e}cnico - IST, Universidade de Lisboa, Portugal}
\newcommand{\Idaho}{Idaho State University, Pocatello, ID 83209, USA}
\newcommand{\Illinoisinstitute}{Illinois Institute of Technology, Chicago, IL 60616, USA}
\newcommand{\Imperial}{Imperial College of Science Technology and Medicine, London SW7 2BZ, United Kingdom}
\newcommand{\IndGuwahati}{Indian Institute of Technology Guwahati, Guwahati, 781 039, India}
\newcommand{\IndHyderabad}{Indian Institute of Technology Hyderabad, Hyderabad, 502285, India}
\newcommand{\Indiana}{Indiana University, Bloomington, IN 47405, USA}
\newcommand{\Ingenieria}{Universidad Nacional de Ingenier{\'\i}a, Lima 25, Per{\'u}}
\newcommand{\Iowa}{University of Iowa, Iowa City, IA 52242, USA}
\newcommand{\IowaState}{Iowa State University, Ames, Iowa 50011, USA}
\newcommand{\Iwate}{Iwate University, Morioka, Iwate 020-8551, Japan}
\newcommand{\Jammu}{University of Jammu, Jammu-180006, India}
\newcommand{\Jawaharlal}{Jawaharlal Nehru University, New Delhi 110067, India}
\newcommand{\Jeonbuk}{Jeonbuk National University, Jeonrabuk-do 54896, South Korea}
\newcommand{\Jyvaskyla}{University of Jyvaskyla, FI-40014, Finland}
\newcommand{\KEK}{High Energy Accelerator Research Organization (KEK), Ibaraki, 305-0801, Japan}
\newcommand{\KISTI}{Korea Institute of Science and Technology Information, Daejeon, 34141, South Korea}
\newcommand{\KL}{K L University, Vaddeswaram, Andhra Pradesh 522502, India}
\newcommand{\Kansasstate}{Kansas State University, Manhattan, KS 66506, USA}
\newcommand{\Kavli}{Kavli Institute for the Physics and Mathematics of the Universe, Kashiwa, Chiba 277-8583, Japan}
\newcommand{\Kure}{National Institute of Technology, Kure College, Hiroshima, 737-8506, Japan}
\newcommand{\Kyiv}{Kyiv National University, 01601 Kyiv, Ukraine}
\newcommand{\LIP}{Laborat{\'o}rio de Instrumenta{\c{c}}{\~a}o e F{\'\i}sica Experimental de Part{\'\i}culas, 1649-003 Lisboa and 3004-516 Coimbra, Portugal}
\newcommand{\Lal}{Laboratoire de l'Acc{\'e}l{\'e}rateur Lin{\'e}aire, 91440 Orsay, France}
\newcommand{\Lancaster}{Lancaster University, Lancaster LA1 4YB, United Kingdom}
\newcommand{\LawrenceBerkeley}{Lawrence Berkeley National Laboratory, Berkeley, CA 94720, USA}
\newcommand{\Liverpool}{University of Liverpool, L69 7ZE, Liverpool, United Kingdom}
\newcommand{\LosAlmos}{Los Alamos National Laboratory, Los Alamos, NM 87545, USA}
\newcommand{\Louisanastate}{Louisiana State University, Baton Rouge, LA 70803, USA}
\newcommand{\Lucknow}{University of Lucknow, Uttar Pradesh 226007, India}
\newcommand{\Madrid}{Madrid Autonoma University and IFT UAM/CSIC, 28049 Madrid, Spain}
\newcommand{\Manchester}{University of Manchester, Manchester M13 9PL, United Kingdom}
\newcommand{\Massinsttech}{Massachusetts Institute of Technology, Cambridge, MA 02139, USA}
\newcommand{\Michigan}{University of Michigan, Ann Arbor, MI 48109, USA}
\newcommand{\Michiganstate}{Michigan State University, East Lansing, MI 48824, USA}
\newcommand{\MilanoBicocca}{Universit{\`a} del Milano-Bicocca, 20126 Milano, Italy}
\newcommand{\MilanoUniv}{Universit{\`a} degli Studi di Milano, I-20133 Milano, Italy}
\newcommand{\Minnduluth}{University of Minnesota Duluth, Duluth, MN 55812, USA}
\newcommand{\Minntwin}{University of Minnesota Twin Cities, Minneapolis, MN 55455, USA}
\newcommand{\Mississippi}{University of Mississippi, University, MS 38677 USA}
\newcommand{\Newmexico}{University of New Mexico, Albuquerque, NM 87131, USA}
\newcommand{\Niewodniczanski}{H. Niewodnicza{\'n}ski Institute of Nuclear Physics, Polish Academy of Sciences, Cracow, Poland}
\newcommand{\Nikhef}{Nikhef National Institute of Subatomic Physics, 1098 XG Amsterdam, Netherlands}
\newcommand{\Northdakota}{University of North Dakota, Grand Forks, ND 58202-8357, USA}
\newcommand{\Northernillinois}{Northern Illinois University, DeKalb, Illinois 60115, USA}
\newcommand{\Northwestern}{Northwestern University, Evanston, Il 60208, USA}
\newcommand{\NotreDame}{University of Notre Dame, Notre Dame, IN 46556, USA}
\newcommand{\Ohiostate}{Ohio State University, Columbus, OH 43210, USA}
\newcommand{\OregonState}{Oregon State University, Corvallis, OR 97331, USA}
\newcommand{\Oxford}{University of Oxford, Oxford, OX1 3RH, United Kingdom}
\newcommand{\PacificNorthwest}{Pacific Northwest National Laboratory, Richland, WA 99352, USA}
\newcommand{\Padova}{Universt{\`a} degli Studi di Padova, I-35131 Padova, Italy}
\newcommand{\Parisuniversite}{Universit{\'e} de Paris, CNRS, Astroparticule et Cosmologie, F-75006, Paris, France}
\newcommand{\Pavia}{Universit{\`a} degli Studi di Pavia, 27100 Pavia PV, Italy}
\newcommand{\Penn}{University of Pennsylvania, Philadelphia, PA 19104, USA}
\newcommand{\PennState}{Pennsylvania State University, University Park, PA 16802, USA}
\newcommand{\PhysicalResearchLaboratory}{Physical Research Laboratory, Ahmedabad 380 009, India}
\newcommand{\Pisa}{Universit{\`a} di Pisa, I-56127 Pisa, Italy}
\newcommand{\Pitt}{University of Pittsburgh, Pittsburgh, PA 15260, USA}
\newcommand{\Pontificia}{Pontificia Universidad Cat{\'o}lica del Per{\'u}, Lima, Per{\'u}}
\newcommand{\PuertoRico}{University of Puerto Rico, Mayaguez 00681, Puerto Rico, USA}
\newcommand{\Punjab}{Punjab Agricultural University, Ludhiana 141004, India}
\newcommand{\Radboud}{Radboud University, NL-6525 AJ Nijmegen, Netherlands}
\newcommand{\Rochester}{University of Rochester, Rochester, NY 14627, USA}
\newcommand{\Royalholloway}{Royal Holloway College London, TW20 0EX, United Kingdom}
\newcommand{\Rutgers}{Rutgers University, Piscataway, NJ, 08854, USA}
\newcommand{\Rutherford}{STFC Rutherford Appleton Laboratory, Didcot OX11 0QX, United Kingdom}
\newcommand{\SLAC}{SLAC National Accelerator Laboratory, Menlo Park, CA 94025, USA}
\newcommand{\SURF}{Sanford Underground Research Facility, Lead, SD, 57754, USA}
\newcommand{\Salento}{Universit{\`a} del Salento, 73100 Lecce, Italy}
\newcommand{\SergioArboleda}{Universidad Sergio Arboleda, 11022 Bogot{\'a}, Colombia}
\newcommand{\Sheffield}{University of Sheffield, Sheffield S3 7RH, United Kingdom}
\newcommand{\SouthDakotaSchool}{South Dakota School of Mines and Technology, Rapid City, SD 57701, USA}
\newcommand{\SouthDakotaState}{South Dakota State University, Brookings, SD 57007, USA}
\newcommand{\Southcarolina}{University of South Carolina, Columbia, SC 29208, USA}
\newcommand{\SouthernMethodist}{Southern Methodist University, Dallas, TX 75275, USA}
\newcommand{\StonyBrook}{Stony Brook University, SUNY, Stony Brook, New York 11794, USA}
\newcommand{\Sussex}{University of Sussex, Brighton, BN1 9RH, United Kingdom}
\newcommand{\Syracuse}{Syracuse University, Syracuse, NY 13244, USA}
\newcommand{\Tennknox}{University of Tennessee at Knoxville, TN, 37996, USA}
\newcommand{\TexasAM}{Texas A{\&}M University - Corpus Christi, Corpus Christi, TX 78412, USA}
\newcommand{\TexasArlington}{University of Texas at Arlington, Arlington, TX 76019, USA}
\newcommand{\Texasaustin}{University of Texas at Austin, Austin, TX 78712, USA}
\newcommand{\Toronto}{University of Toronto, Toronto, Ontario M5S 1A1, Canada}
\newcommand{\Tufts}{Tufts University, Medford, MA 02155, USA}
\newcommand{\UCIII}{Universidad Carlos III de Madrid, Av. de la Universidad, 30, 28911 Madrid, Spain}
\newcommand{\Unifesp}{Universidade Federal de S{\~a}o Paulo, 09913-030, S{\~a}o Paulo, Brazil}
\newcommand{\UniversityCollegeLondon}{University College London, London, WC1E 6BT, United Kingdom}
\newcommand{\ValleyCity}{Valley City State University, Valley City, ND 58072, USA}
\newcommand{\VariableEnergy}{Variable Energy Cyclotron Centre, 700 064 West Bengal, India}
\newcommand{\VirginiaTech}{Virginia Tech, Blacksburg, VA 24060, USA}
\newcommand{\Warsaw}{University of Warsaw, 00-927 Warsaw, Poland}
\newcommand{\Warwick}{University of Warwick, Coventry CV4 7AL, United Kingdom}
\newcommand{\Wichita}{Wichita State University, Wichita, KS 67260, USA}
\newcommand{\WilliamMary}{William and Mary, Williamsburg, VA 23187, USA}
\newcommand{\Wisconsin}{University of Wisconsin Madison, Madison, WI 53706, USA}
\newcommand{\Yale}{Yale University, New Haven, CT 06520, USA}
\newcommand{\Yerevan}{Yerevan Institute for Theoretical Physics and Modeling, Yerevan 0036, Armenia}
\newcommand{\York}{York University, Toronto M3J 1P3, Canada}
\affiliation{\Amsterdam}
\affiliation{\Antananarivo}
\affiliation{\AntonioNarino}
\affiliation{\Argonne}
\affiliation{\Arizona}
\affiliation{\Asuncion}
\affiliation{\Athens}
\affiliation{\Atlantico}
\affiliation{\Banaras}
\affiliation{\Basel}
\affiliation{\Bern}
\affiliation{\Beykent}
\affiliation{\Birmingham}
\affiliation{\BolognaUniversity}
\affiliation{\Boston}
\affiliation{\Bristol}
\affiliation{\Brookhaven}
\affiliation{\Bucharest}
\affiliation{\CBPF}
\affiliation{\CEASaclay}
\affiliation{\CERN}
\affiliation{\CIEMAT}
\affiliation{\CUSB}
\affiliation{\CalBerkeley}
\affiliation{\CalDavis}
\affiliation{\CalIrvine}
\affiliation{\CalLosangeles}
\affiliation{\CalRiverside}
\affiliation{\CalSantabarbara}
\affiliation{\Caltech}
\affiliation{\Cambridge}
\affiliation{\Campinas}
\affiliation{\CataniaUniversitadi}
\affiliation{\Charles}
\affiliation{\Chicago}
\affiliation{\ChungAng}
\affiliation{\Cincinnati}
\affiliation{\Cinvestav}
\affiliation{\Colima}
\affiliation{\ColoradoBoulder}
\affiliation{\ColoradoState}
\affiliation{\Columbia}
\affiliation{\CzechAcademyofSciences}
\affiliation{\CzechTechnical}
\affiliation{\DakotaState}
\affiliation{\Dallas}
\affiliation{\DannecyleVieux}
\affiliation{\Daresbury}
\affiliation{\Drexel}
\affiliation{\Duke}
\affiliation{\Durham}
\affiliation{\EIA}
\affiliation{\ETH}
\affiliation{\Edinburgh}
\affiliation{\FCULport}
\affiliation{\FederaldeAlfenas}
\affiliation{\FederaldeGoias}
\affiliation{\FederaldeSaoCarlos}
\affiliation{\FederaldoABC}
\affiliation{\FederaldoRio}
\affiliation{\Fermi}
\affiliation{\Florida}
\affiliation{\Fluminense}
\affiliation{\Genova}
\affiliation{\Georgian}
\affiliation{\GranSasso}
\affiliation{\GranSassoLab}
\affiliation{\Granada}
\affiliation{\Grenoble}
\affiliation{\Guanajuato}
\affiliation{\Harish}
\affiliation{\Harvard}
\affiliation{\Hawaii}
\affiliation{\Houston}
\affiliation{\Hyderabad}
\affiliation{\IFAE}
\affiliation{\IFIC}
\affiliation{\INFNBologna}
\affiliation{\INFNCatania}
\affiliation{\INFNGenova}
\affiliation{\INFNLecce}
\affiliation{\INFNMilanBicocca}
\affiliation{\INFNMilano}
\affiliation{\INFNNapoli}
\affiliation{\INFNPadova}
\affiliation{\INFNPavia}
\affiliation{\INFNSud}
\affiliation{\INR}
\affiliation{\IPLyon}
\affiliation{\IPM}
\affiliation{\ISTlisboa}
\affiliation{\Idaho}
\affiliation{\Illinoisinstitute}
\affiliation{\Imperial}
\affiliation{\IndGuwahati}
\affiliation{\IndHyderabad}
\affiliation{\Indiana}
\affiliation{\Ingenieria}
\affiliation{\Iowa}
\affiliation{\IowaState}
\affiliation{\Iwate}
\affiliation{\Jammu}
\affiliation{\Jawaharlal}
\affiliation{\Jeonbuk}
\affiliation{\Jyvaskyla}
\affiliation{\KEK}
\affiliation{\KISTI}
\affiliation{\KL}
\affiliation{\Kansasstate}
\affiliation{\Kavli}
\affiliation{\Kure}
\affiliation{\Kyiv}
\affiliation{\LIP}
\affiliation{\Lal}
\affiliation{\Lancaster}
\affiliation{\LawrenceBerkeley}
\affiliation{\Liverpool}
\affiliation{\LosAlmos}
\affiliation{\Louisanastate}
\affiliation{\Lucknow}
\affiliation{\Madrid}
\affiliation{\Manchester}
\affiliation{\Massinsttech}
\affiliation{\Michigan}
\affiliation{\Michiganstate}
\affiliation{\MilanoBicocca}
\affiliation{\MilanoUniv}
\affiliation{\Minnduluth}
\affiliation{\Minntwin}
\affiliation{\Mississippi}
\affiliation{\Newmexico}
\affiliation{\Niewodniczanski}
\affiliation{\Nikhef}
\affiliation{\Northdakota}
\affiliation{\Northernillinois}
\affiliation{\Northwestern}
\affiliation{\NotreDame}
\affiliation{\Ohiostate}
\affiliation{\OregonState}
\affiliation{\Oxford}
\affiliation{\PacificNorthwest}
\affiliation{\Padova}
\affiliation{\Parisuniversite}
\affiliation{\Pavia}
\affiliation{\Penn}
\affiliation{\PennState}
\affiliation{\PhysicalResearchLaboratory}
\affiliation{\Pisa}
\affiliation{\Pitt}
\affiliation{\Pontificia}
\affiliation{\PuertoRico}
\affiliation{\Punjab}
\affiliation{\Radboud}
\affiliation{\Rochester}
\affiliation{\Royalholloway}
\affiliation{\Rutgers}
\affiliation{\Rutherford}
\affiliation{\SLAC}
\affiliation{\SURF}
\affiliation{\Salento}
\affiliation{\SergioArboleda}
\affiliation{\Sheffield}
\affiliation{\SouthDakotaSchool}
\affiliation{\SouthDakotaState}
\affiliation{\Southcarolina}
\affiliation{\SouthernMethodist}
\affiliation{\StonyBrook}
\affiliation{\Sussex}
\affiliation{\Syracuse}
\affiliation{\Tennknox}
\affiliation{\TexasAM}
\affiliation{\TexasArlington}
\affiliation{\Texasaustin}
\affiliation{\Toronto}
\affiliation{\Tufts}
\affiliation{\UCIII}
\affiliation{\Unifesp}
\affiliation{\UniversityCollegeLondon}
\affiliation{\ValleyCity}
\affiliation{\VariableEnergy}
\affiliation{\VirginiaTech}
\affiliation{\Warsaw}
\affiliation{\Warwick}
\affiliation{\Wichita}
\affiliation{\WilliamMary}
\affiliation{\Wisconsin}
\affiliation{\Yale}
\affiliation{\Yerevan}
\affiliation{\York}

\author{B.~Abi} \affiliation{\Oxford}
\author{R.~Acciarri} \affiliation{\Fermi}
\author{M.~A.~Acero} \affiliation{\Atlantico}
\author{G.~Adamov} \affiliation{\Georgian}
\author{D.~Adams} \affiliation{\Brookhaven}
\author{M.~Adinolfi} \affiliation{\Bristol}
\author{Z.~Ahmad} \affiliation{\VariableEnergy}
\author{J.~Ahmed} \affiliation{\Warwick}
\author{T.~Alion} \affiliation{\Sussex}
\author{S.~Alonso Monsalve}\email{E-Mail: saul.alonso.monsalve@cern.ch} \affiliation{\CERN}
% ...

\collaboration{The DUNE Collaboration}

\noaffiliation

\maketitle

\section{Introduction to DUNE}
 Over the last twenty years neutrino oscillations~\cite{MNS1962,Pontecorvo1967} have become well-established~\cite{oscSuperK,oscSNO,oscKamland,oscK2K,oscDayaBay,MINOSFinal3Flav,Acero:2019ksn,Abe:2019vii} and the field is moving into the precision measurement era. The PMNS~\cite{MNS1962,Pontecorvo1967} neutrino oscillation formalism describes observed data with six fundamental parameters. These are three angles describing the rotation between the neutrino mass and flavor eigenstates, two mass splittings (differences between the squared masses of the neutrino mass states), and $CP$-violating phase, $\delta_{CP}$. If $\sin\left(\delta_{CP}\right)$ is non-zero then the vacuum oscillation probabilities of neutrinos and antineutrinos will be different.
 DUNE~\cite{dunetdr} is a next-generation neutrino oscillation experiment with a primary scientific goal of making precise measurements of the parameters governing long-baseline neutrino oscillation. A particular priority is the observation of $CP$-violation in the neutrino sector. In DUNE, a muon neutrino ($\nu_\mu$)- or muon antineutrino ($\bar{\nu}_{\mu}$)-dominated beam will be produced by the Long-Baseline Neutrino Facility (LBNF) beamline and characterized by a near detector (ND) at Fermilab before the neutrinos travel 1285\,km to the Sanford Underground Research Facility (SURF). The far detector (FD) will consist of four 10\,kt (fiducial) liquid argon time projection chamber (LArTPC) detectors. Oscillation probabilities are inferred from comparison of the observed neutrino spectra at the near and far detectors which are used to constrain values of the neutrino oscillation parameters.

\subsection{$CP$-violation measurement}
Symmetries under charge conjugation and parity inversion are both maximally violated by the weak interaction. Their combined operation has been shown to be violated, to a small degree, by quark mixing processes~\cite{parityViolation,chargeParityViolation}. The neutrino oscillation formalism allows for an analogous process in lepton flavor mixing which can be measured with neutrino oscillations. DUNE is sensitive to four neutrino oscillation parameters, namely $ \Delta m^{2}_{31},~\theta_{23},~\theta_{13}~\textrm{and}~\delta_{CP} $, which can be measured using four data samples: two for neutrinos and two for antineutrinos.
Two beam configurations with opposite polarities of the magnetic focusing horns are used to produce these samples: ``forward horn current'' (FHC) mode produces a predominantly $\nu_\mu$ beam while a primarily $\bar{\nu}_\mu$ beam is produced in ``reverse horn current'' (RHC) mode. 
The FD data used in the oscillation analysis measure the ``disappearance'' channels (i.e. $\nu_\mu \rightarrow \nu_\mu$ and $\bar{\nu}_\mu \rightarrow \bar{\nu}_\mu$), which are primarily sensitive to $|\Delta m^{2}_{31}|$ and $\sin^22\theta_{23}$, and the ``appearance'' channels (i.e. $\nu_\mu \rightarrow \nu_e$ and $\bar{\nu}_\mu \rightarrow \bar{\nu}_e$), which are sensitive to all four parameters, including the sign of $\Delta m^{2}_{31}$. In all of these samples, interactions where the neutrinos scatter via charged-current (CC) exchange off the nuclei in the far detector are selected. In a CC interaction, the final state includes a charged lepton with the same flavor as the incoming neutrino and one or more hadrons, depending on the details of the interaction. Therefore, a critical aspect of event selection is the ability to identify the flavor of the final-state lepton. Thus it is key to be able to efficiently identify the signal (i.e. CC $\nu_\mu$, CC $\bar{\nu}_\mu$, CC $\nu_e$ and CC $\bar{\nu}_e$) interactions and have a powerful rejection of background events. At the energies relevant to the DUNE oscillation analysis, a final-state muon produces a long, straight track in the detector, while a final-state electron produces an electromagnetic (EM) shower. Examples of signal CC $\nu_e$ and CC $\nu_\mu$ interactions are shown in Figs.~\ref{fig:views} and~\ref{fig:view0_nc}, respectively.

\begin{figure}[htb] 
	\begin{subfigure}{0.45\textwidth}
		\includegraphics[width=\linewidth]{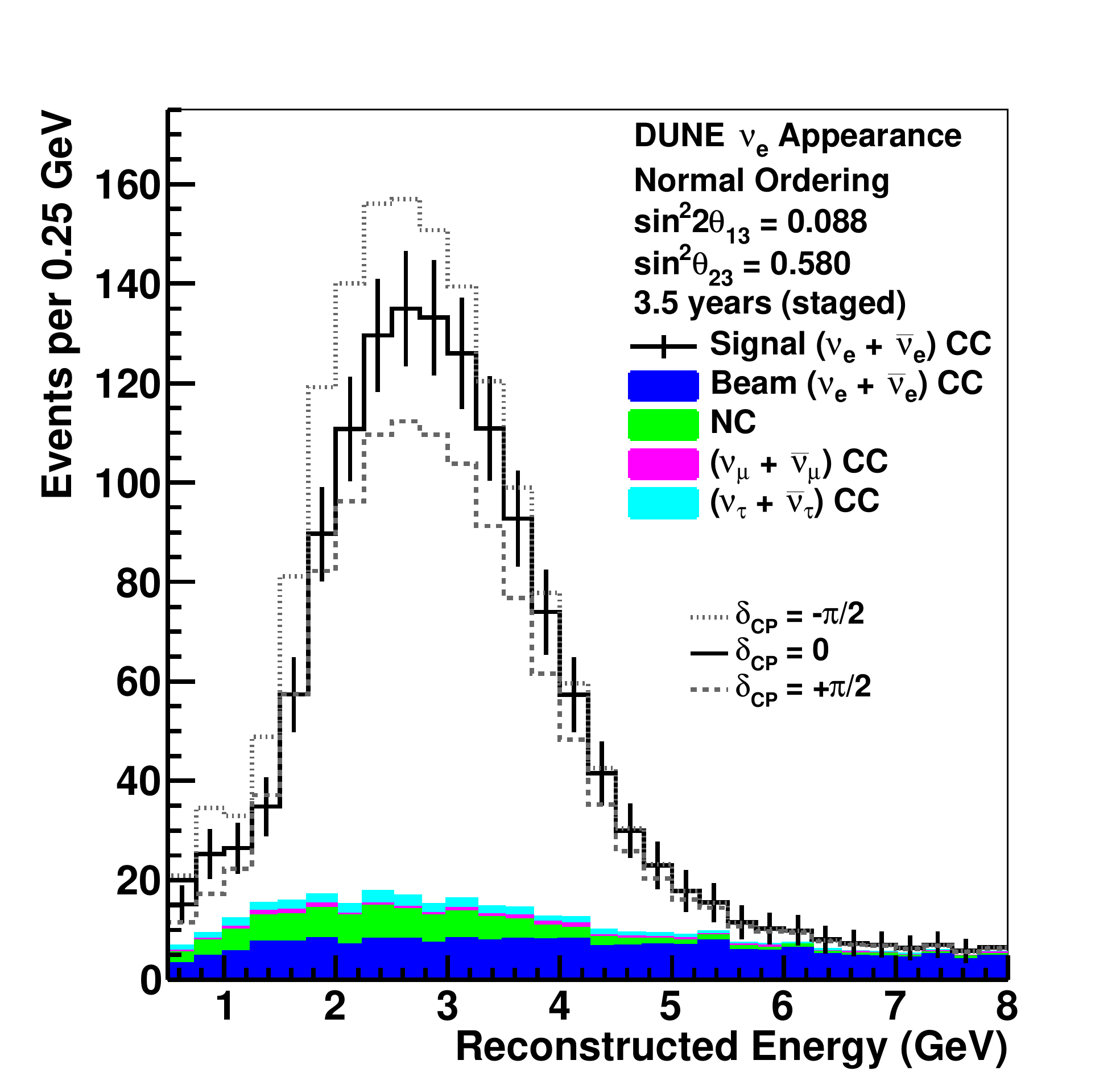}
		\caption{Neutrino mode.} 
	\end{subfigure}
	\hspace*{\fill} 
	\begin{subfigure}{0.45\textwidth}
		\includegraphics[width=\linewidth]{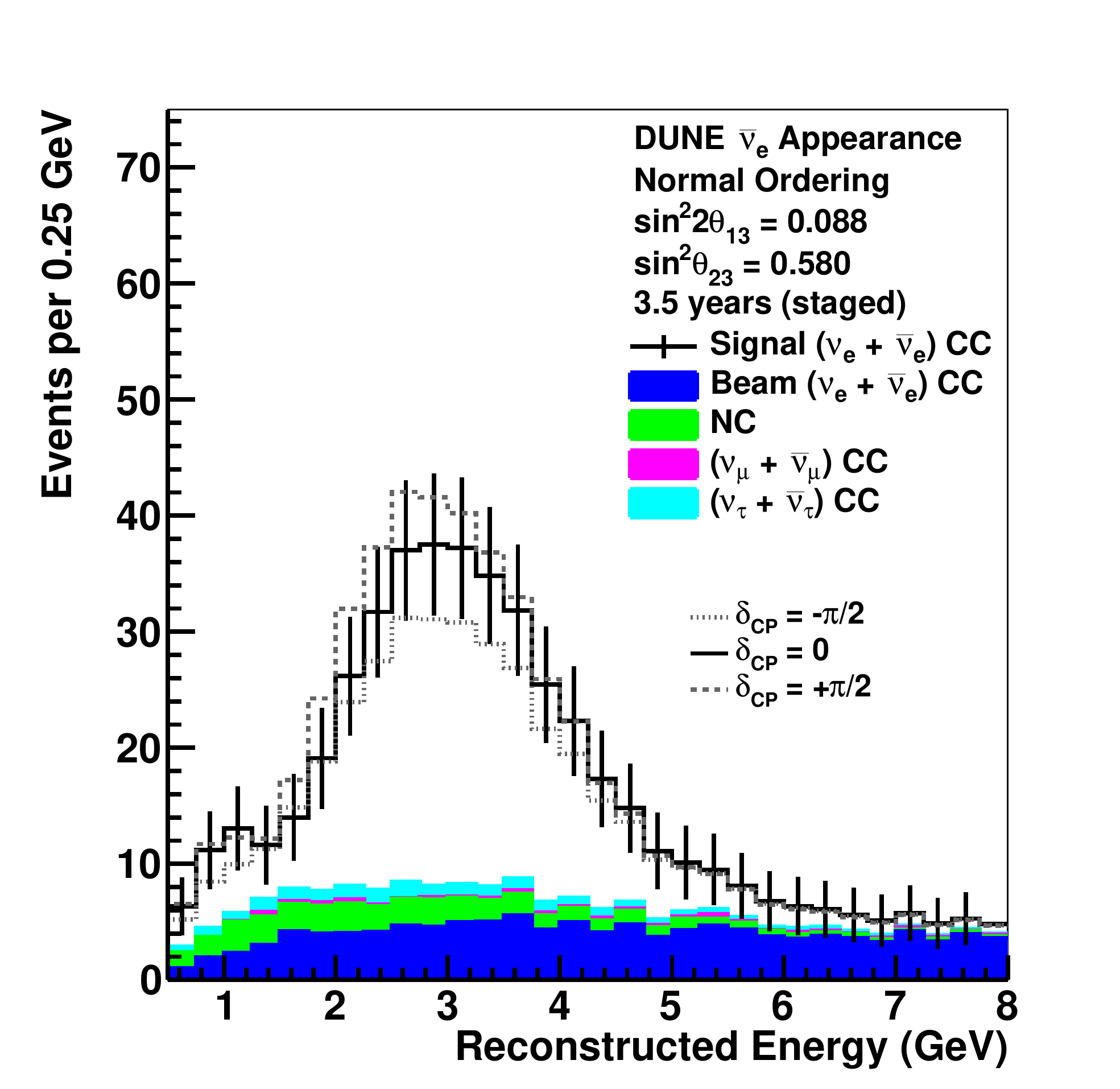}
		\caption{Antineutrino mode.} 
	\end{subfigure}
	\hspace*{\fill} 
	\caption{Reconstructed energy distribution of $\nu_e$ and $\bar{\nu}_e$ CC-like events selected by the convolutional neural network algorithm (CVN) assuming 3.5 years (staged) running in the neutrino-beam mode (a) and antineutrino-beam mode (b), for a total of seven years (staged) exposure.  The plots assume normal mass ordering and include curves for $\delta_{CP}$ = -$\pi$/2, 0, and $\pi/2$. Background from $\nu_{\mu}$-CC, $\nu_{\tau}$-CC, intrinsic $\nu_{e}$-CC, and NC interactions are shown as stacked, filled histograms. Figure reproduced from Ref.~\cite{lblPaper}.}
	\label{fig:spectra}
\end{figure}

The main background to the CC $\nu_\mu$ and CC $\bar{\nu}_\mu$ event selections are neutral current (NC) interactions with charged pions ($\pi^\pm$) in the final state that can mimic the $\mu^\pm$, an example of which is shown in Fig.~\ref{fig:view1_nc}. Neutral current interactions with a final-state $\pi^0$ meson, such as the one shown in Fig.~\ref{fig:view2_nc}, where the photons from $\pi^0$ decay may mimic the EM shower from an electron, form the primary reducible background to the CC $\nu_e$ and CC $\bar{\nu}_e$ event selections. A small fraction of electron neutrinos are intrinsic to the beam (and thus are not the result of neutrino oscillations). These events form a background for the oscillation analysis as they are indistinguishable from CC $\nu_e$ appearance events. Once the four samples have been selected and the neutrino energy has been reconstructed, a fit is performed to the reconstructed neutrino energy distributions in the four samples to extract the neutrino oscillation parameters $\theta_{13}$, $\theta_{23}$, $\Delta m^{2}_{31}$, and $\delta_{CP}$. This fit accounts for the effects of systematic uncertainties, including the constraints on those uncertainties from fits to ND data. Figure~\ref{fig:spectra} shows the appearance samples and how they are expected to vary with the true value of $\delta_{CP}$, for a data collection period of 3.5\,years staged running in both FHC and RHC beam modes. The staging plan assumes two FD modules are ready at the start of the beam data taking, and modules three and four become operational after one year and two years, respectively. Full details of the DUNE staging plan and the oscillation analysis, including the assumed oscillation parameters, are provided in Ref.~\cite{lblPaper}.

\subsection{DUNE Far Detector}
Neutrinos are detected via their interaction products i.e. observation of the leptons and hadrons that are produced when the neutrinos interact in the detector. In the single-phase LArTPC design that will be used for the first DUNE FD module, three wire readout planes collect the ionization charge that is generated when charged particles traverse the liquid argon volume. The ionization charge drifts in a constant electric field to the readout planes and the drift time provides a third dimension of position information, giving rise to the name ``time projection chamber.'' The position of the charge observed in each of the three planes is combined with the drift time to create three views of each neutrino interaction. The wires that form the planes are separated by approximately 5\,mm giving the FD a fine-grained sampling of the neutrino interaction products. The electronic signals from the wires are sampled at a rate of 2\,MHz, giving a similar effective spatial resolution in the time direction. Two of the wire planes are induction planes, biased to be transparent to the drifting electrons, such that they induce net-zero fluctuation in the wire current as they pass the wire plane. The third view is called the collection plane as it actually collects the drifting electrons. The four DUNE FD modules may not all have identical designs, but they will all produce similar images of the neutrino interactions, so the performance of the single-phase design is used throughout this article. Other potential designs must have at least the same sampling capabilities as the single-phase design, if not better, to be considered.

\subsection{DUNE Simulation and Reconstruction}\label{sec:reco}
Neutrino interactions in the far detector are simulated within the LArSoft~\cite{Church:2013hea} framework, using the neutrino flux from a GEANT4-based~\cite{Agostinelli:2002hh} simulation of the LBNF beamline, the GENIE~\cite{Alam:2015nkk} neutrino interaction generator (version 2.12.10), and a GEANT4-based (version 10.3.01) detector simulation. Detector response to, and readout of, the ionization charge is also simulated in LArSoft. Raw detector waveforms are processed to remove the impact of the electric field and electronics response; this process is referred to as ``deconvolution'' and the resulting deconvolved waveforms contain calibrated charge information. Current fluctuations in the wires above threshold, or ``hits'', are parameterized by Gaussian functions fit to deconvolved waveforms around local maxima. A reconstruction algorithm is used to cluster hits linked in space and time into groups associated with a particular physical object, such as a track or shower.
More details of the DUNE simulation and reconstruction are available in Ref.~\cite{dunetdr}.

The energy of the incoming neutrino in CC events is estimated by a dedicated algorithm that adds the reconstructed lepton and hadronic energies, using particles reconstructed by Pandora~\cite{pandora,pandorauboone}. Pandora uses a multi-algorithm approach to reconstruct all the visible particles produced in neutrino interactions. It provides a hierarchy of reconstructed particles, representing particles produced at the interaction vertex and their decays or subsequent interactions. If the event is selected as CC $\nu_{\mu}$, the neutrino energy is estimated as the sum of the energy of the longest reconstructed track and the hadronic energy, where the energy of the longest reconstructed track is estimated from its range if the track is contained in the detector and from multiple Coulomb scattering if the track exits the detector. The hadronic energy is estimated from the energy associated with reconstructed hits that are not in the longest track. If the event is selected as CC $\nu_e$, the energy of the neutrino is estimated as the sum of the energy of the reconstructed shower with the highest energy and the hadronic energy. In all cases, simulation-based corrections for missing energy (due to undetected particles, reconstruction errors, etc) are applied.

\section{CVN neutrino interaction classifier}

The DUNE Convolutional Visual Network (CVN) classifies neutrino interactions in the DUNE FD through image recognition techniques. In general terms it is a convolutional neural network (CNN)~\cite{firstCNN}. The main feature of CNNs is that they apply a series of filters (using convolutions, hence the name of the CNN) to the images to extract features that allow the CNN to classify the images~\cite{CNNClassification}. Each of the filters - also known as kernels - consists of a set of values that are learnt by the CNN through the training process. CNNs are typically deep neural networks that consist of many convolutional layers, with the output from one convolutional layer forming the input to the next. Similar techniques have been demonstrated to outperform traditional event reconstruction-based methods to classify neutrino interactions~\cite{novacvn,microboonecvn}.

Convolutional neural networks make use of learned kernel operations, usually followed by spatial pooling, applied in sequence to extract increasingly powerful and abstract features. In domains such as natural image analysis where important features of the data are locally spatially correlated they now greatly outperform previous state-of-the-art techniques that relied on manual feature extraction and simpler Machine Learning methods~\cite{representationLearning,Goodfellow-et-al-2016-deep,advancesCNNs,CNNApplications}. Recently they have proven to also be appropriate for the analysis of signals in particle physics detectors~\cite{CNNatLHC,Jet,Jet2}. They have found particular success in neutrino experiments where signals can arrive at any location in large uniform detector volumes~\cite{radovicNature,novacvn,microboonecvn, Adams:2019uqx}, and the characteristic translational invariance of CNN methods represents an advantage rather than a challenge.

\subsection{Inputs to the CVN}
\label{sec:inputs}
Figure~\ref{fig:cvnarchitecture} shows that there are three inputs to the CVN. The three inputs are 500$\times$500 pixel images of simulated neutrino interactions with one image produced for each of the three readout views of the LArTPC. The images are produced at the hit-level stage of the reconstruction algorithms and are hence independent of any potential errors in high-level reconstruction such as clustering, track-finding and shower reconstruction. The images are produced in (wire number, time) coordinates, where the wire number is simply the wire on which the reconstructed hit was detected, and the time is the interval from when the interaction happened to when the hit was detected on that wire (given by the peak time of the hit). The color of the pixel gives the hit charge where white shows that no hit was recorded for that pixel. Each pixel represents approximately 5\,mm in the wire coordinate due to the spatial separation of the wires in the readout plane, and the time coordinate is down-sampled to approximately correspond to the same 5\,mm size after consideration of the electron drift velocity within the LArTPC.

Convolutional neural networks operate on fixed-size images, hence the neutrino interaction images must all be of a fixed size. To facilitate this, interactions that span more than 500 wires in a given view are cropped to fit in $500\times500$ pixel images. The steps below are used to find the 500 pixels in the wire coordinate:
\begin{enumerate}
    \item Integrate the charge on each wire.
    \item Scan from low wire number, where low wire number corresponds to the upstream end of the detector, to high wire number and check the following 20 wires for recorded signals. If fewer than five of the 20 subsequent wires have no signals then this wire is chosen as the first column of the image.
    \item  If no wire satisfies the requirement in Step 2, choose the continuous 500 wire range that contains the most deposited charge.
\end{enumerate}
For the time axis, a window of 3200$\,\mu$s centred on the mean time of the hits is formed and divided into 500 bins that fill the 500 pixels. As such, no analogous region-of-interest search is performed.

\begin{figure}[htb] 
	\begin{subfigure}{0.3\textwidth}
		\fbox{\includegraphics[width=0.95\linewidth]{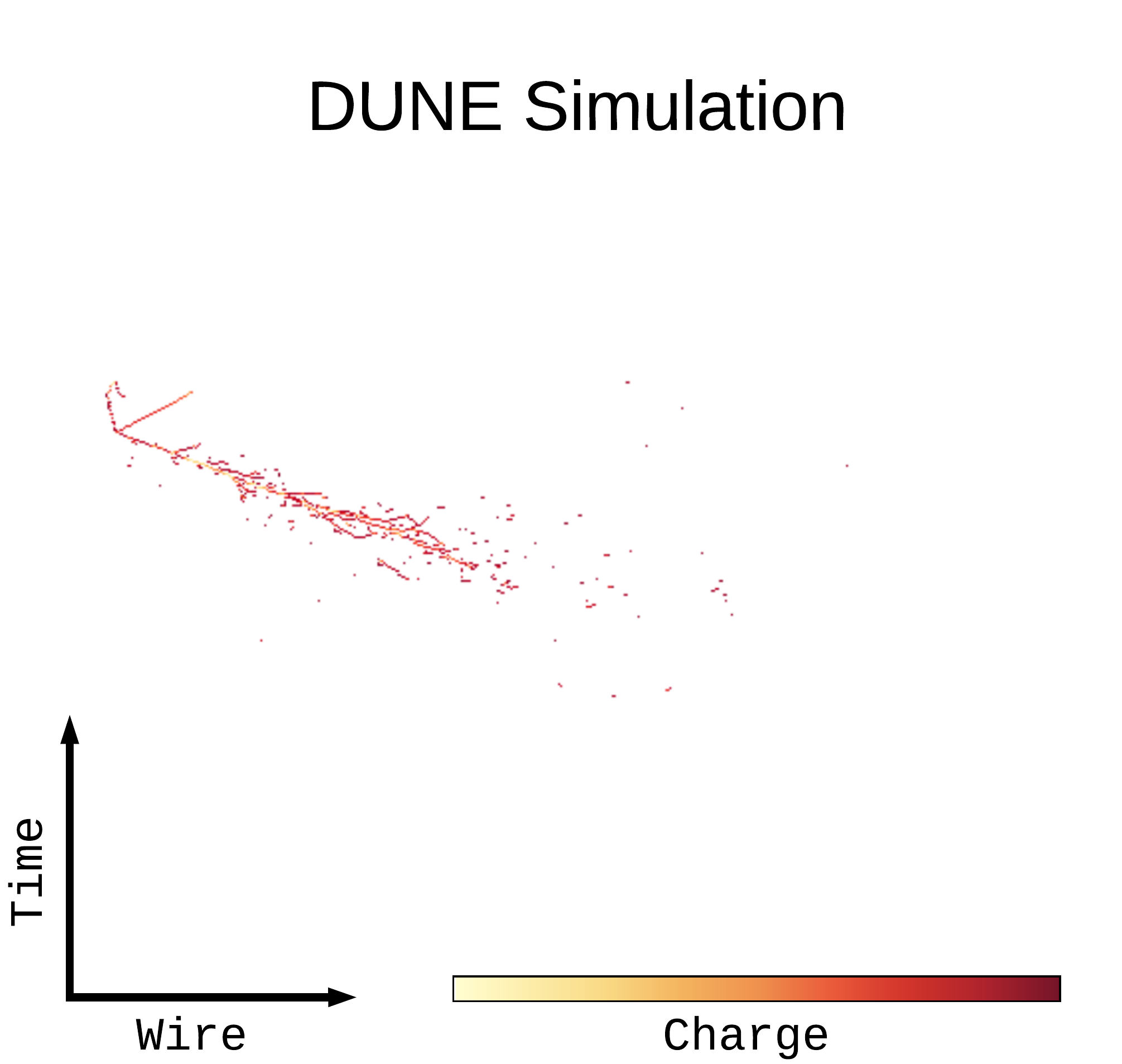}}
		\caption{View 0: Induction Plane.} 
		\label{fig:view0}
	\end{subfigure}
	\hspace*{\fill} 
	\begin{subfigure}{0.3\textwidth}
		\fbox{\includegraphics[width=0.95\linewidth]{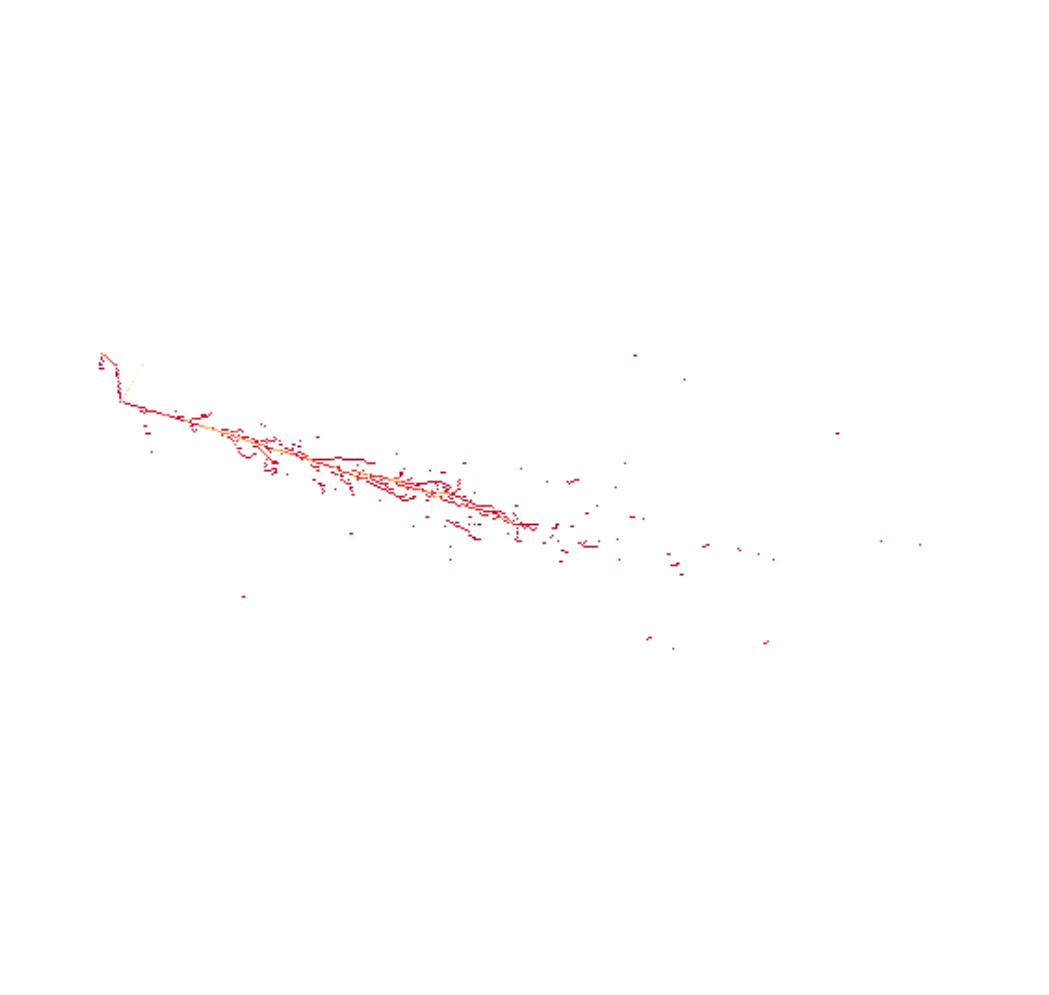}}
		\caption{View 1: Induction Plane.} 
		\label{fig:view1}
	\end{subfigure}
	\hspace*{\fill} 
	\begin{subfigure}{0.3\textwidth}
		\fbox{\includegraphics[width=0.95\linewidth]{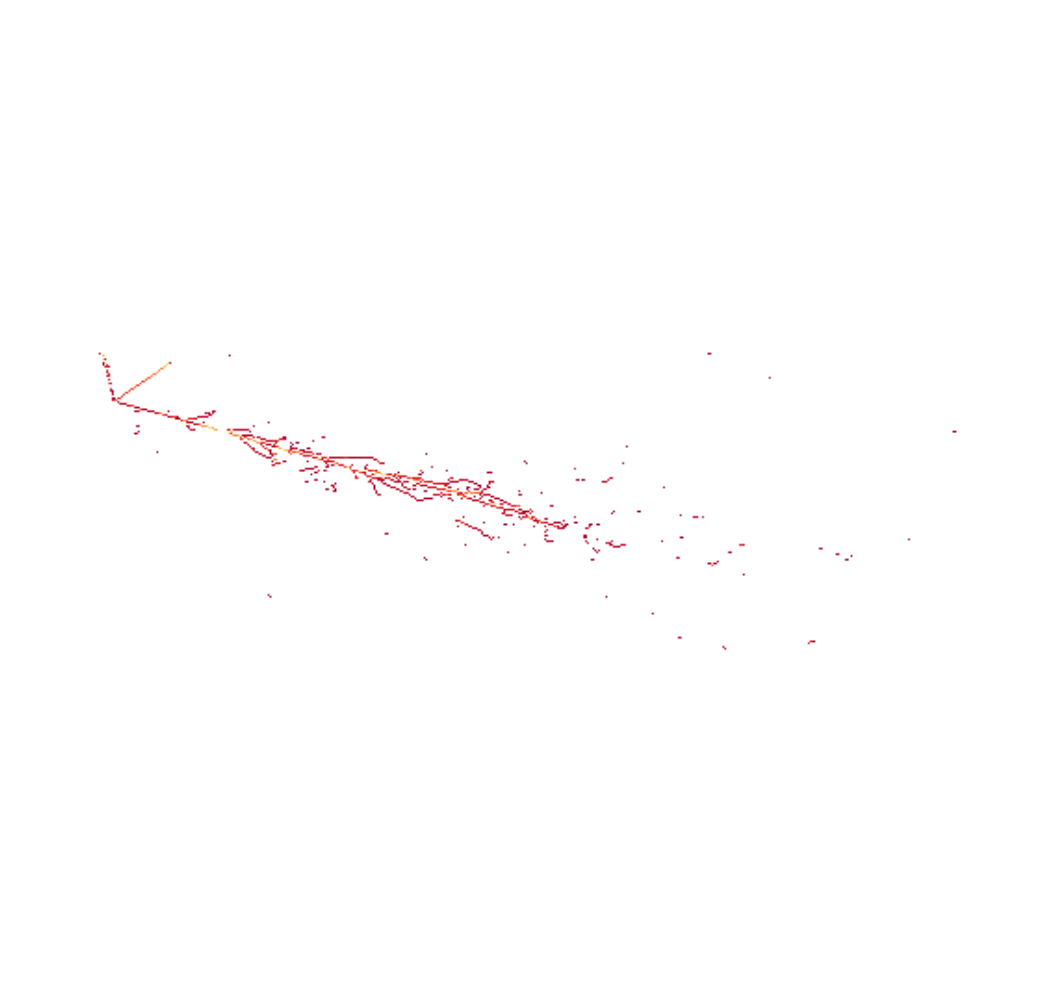}}
		\caption{View 2: Collection Plane.} 
		\label{fig:view2}
	\end{subfigure}
	\caption{A 2.2\,GeV CC $\nu_e$ interaction shown in the three readout views of the DUNE LArTPCs showing the characteristic electromagnetic shower topology. The horizontal axis shows the wire number of the readout plane and the vertical axis shows time. The color scale shows the charge of the energy deposits on the wires.}
	\label{fig:views}
\end{figure}

\begin{figure}[htb] 
	\begin{subfigure}{0.3\textwidth}
		\fbox{\includegraphics[width=0.95\linewidth]{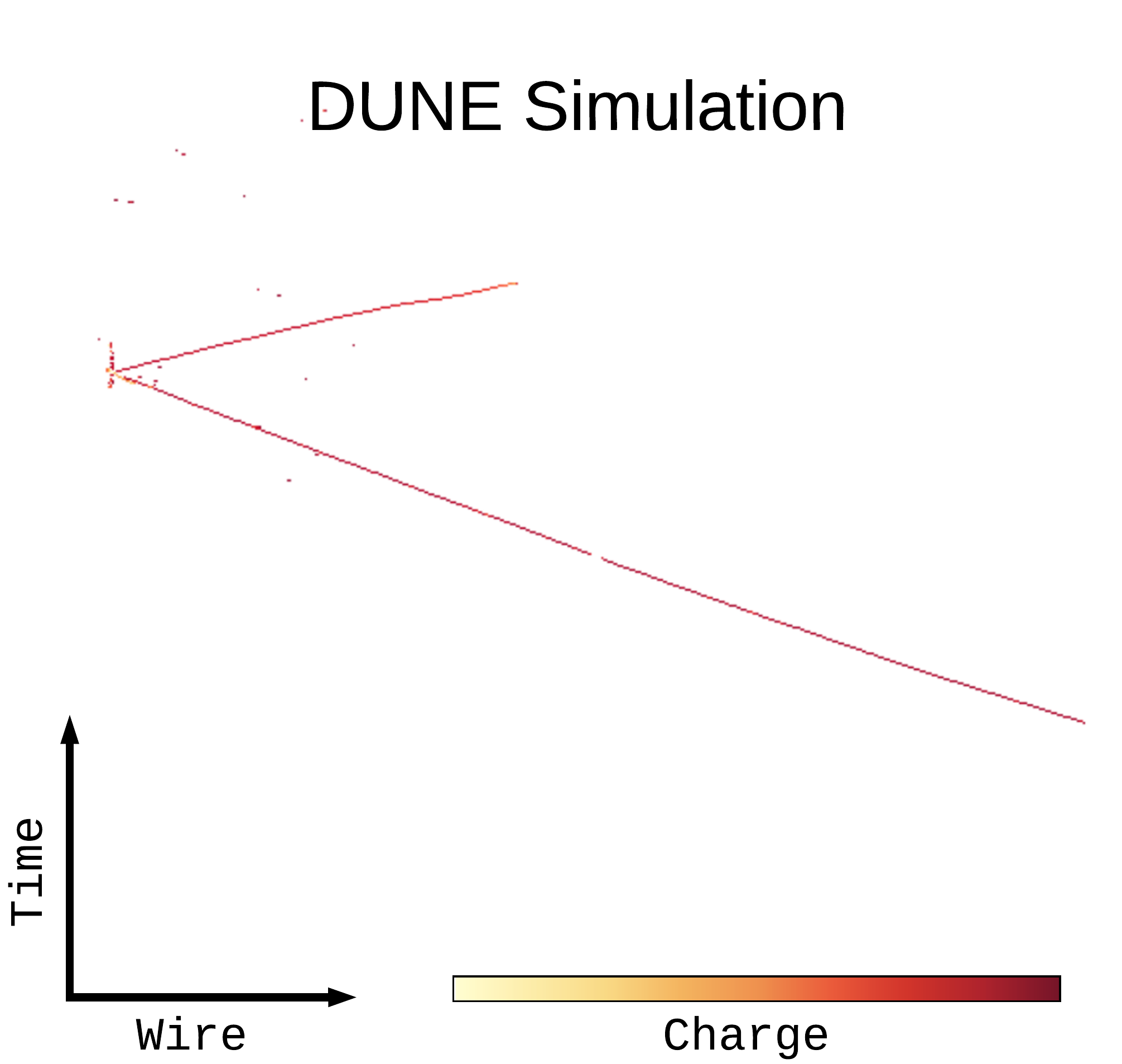}}
		\caption{1.6\,GeV CC $\nu_\mu$.} 
		\label{fig:view0_nc}
	\end{subfigure}
	\hspace*{\fill} 
	\begin{subfigure}{0.3\textwidth}
		\fbox{\includegraphics[width=0.95\linewidth]{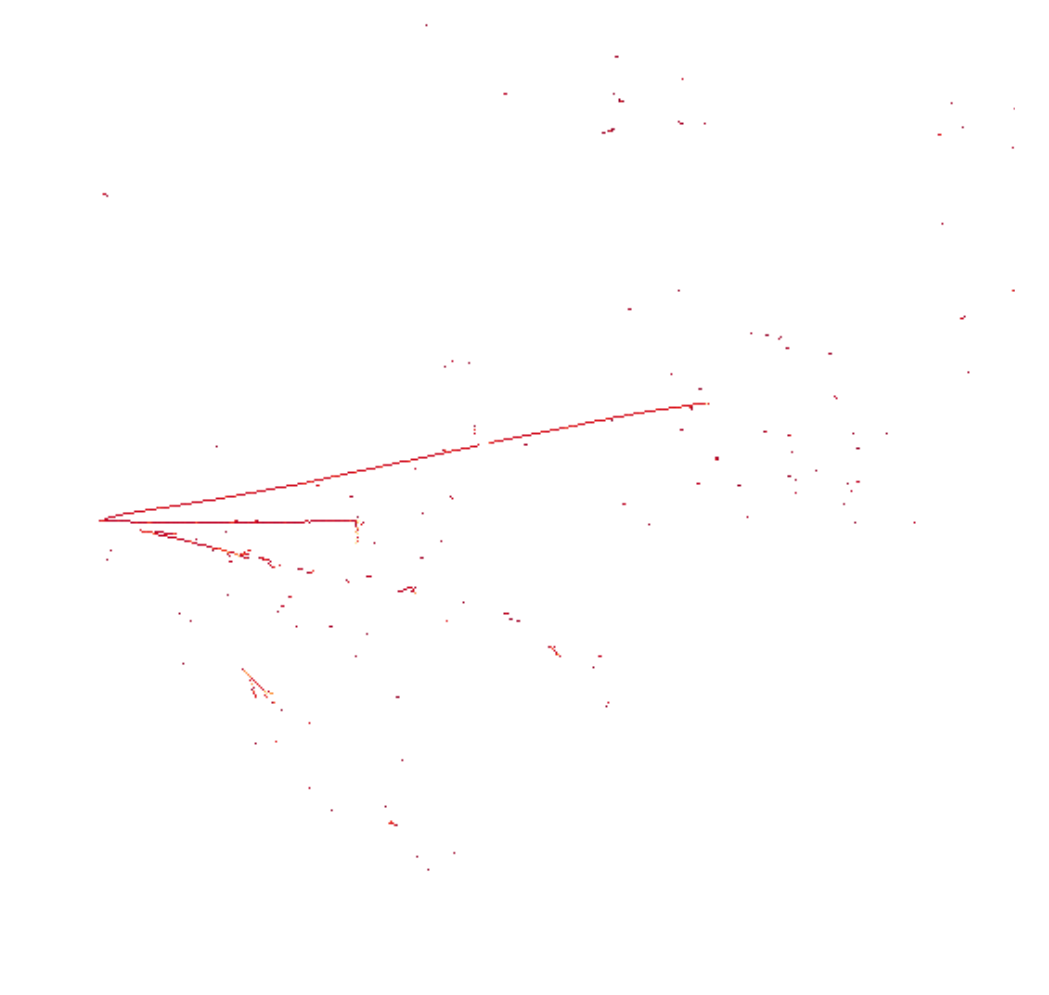}}
		\caption{2.2\,GeV NC 1$\pi^+$.} 
		\label{fig:view1_nc}
	\end{subfigure}
	\hspace*{\fill} 
	\begin{subfigure}{0.3\textwidth}
		\fbox{\includegraphics[width=0.95\linewidth]{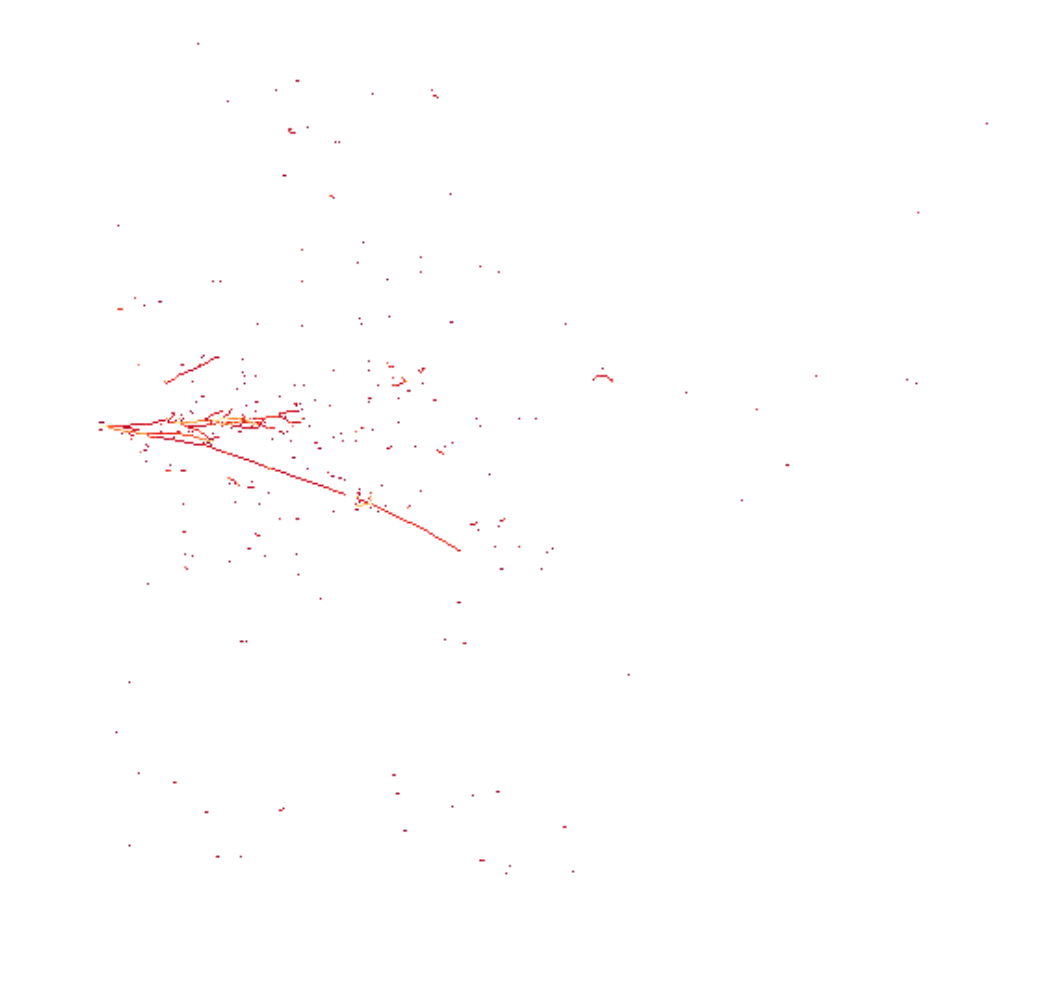}}
		\caption{2.4\,GeV NC 1$\pi^0$.} 
		\label{fig:view2_nc}
	\end{subfigure}
	\caption{Three interactions shown in the collection view: a) a signal CC $\nu_\mu$ interaction, b) an NC interaction with a long $\pi^+$ track and c) an NC interaction with one $\pi^0$. The NC interactions shown in b) and c) form the primary backgrounds to CC $\nu_\mu$ and CC $\nu_e$ event identification, respectively.
	}
	\label{fig:views_numu}
\end{figure}

In order to ensure high quality images of the interactions, images were only produced for events that have their true neutrino interaction vertex within the detector fiducial volume described in Ref.~\cite{lblPaper}. Once the images have been produced, any events that contain any view with fewer than 10 non-zero pixels are removed in order to discount empty and almost empty images from the training and testing datasets. Figure~\ref{fig:views} shows a signal CC $\nu_e$ event as seen in the three detector readout views. Figure~\ref{fig:view0_nc} shows a signal CC $\nu_\mu$ interaction, and example NC background images containing a long $\pi^\pm$ track and a $\pi^0$ are given in Figs.~\ref{fig:view1_nc} and ~\ref{fig:view2_nc}, respectively.

The number of pixels in the images was chosen to maximize the size of the image whilst ensuring that the memory usage during training and inference of the network was manageable. The spatial dimension of the images covers 2.5\,m, meaning any tracks with projected lengths in the readout planes above 2.5\,m will not be fully contained within the image, as is the case for the majority of muon tracks, including the one shown in Fig.~\ref{fig:view0_nc}. However, the key details for the neutrino interaction classification come from the region surrounding the vertex, so this choice of image size does not significantly impact the classification performance.

\subsection{Network architecture}
\begin{figure}
    \centering
		\includegraphics[width=0.8\linewidth]{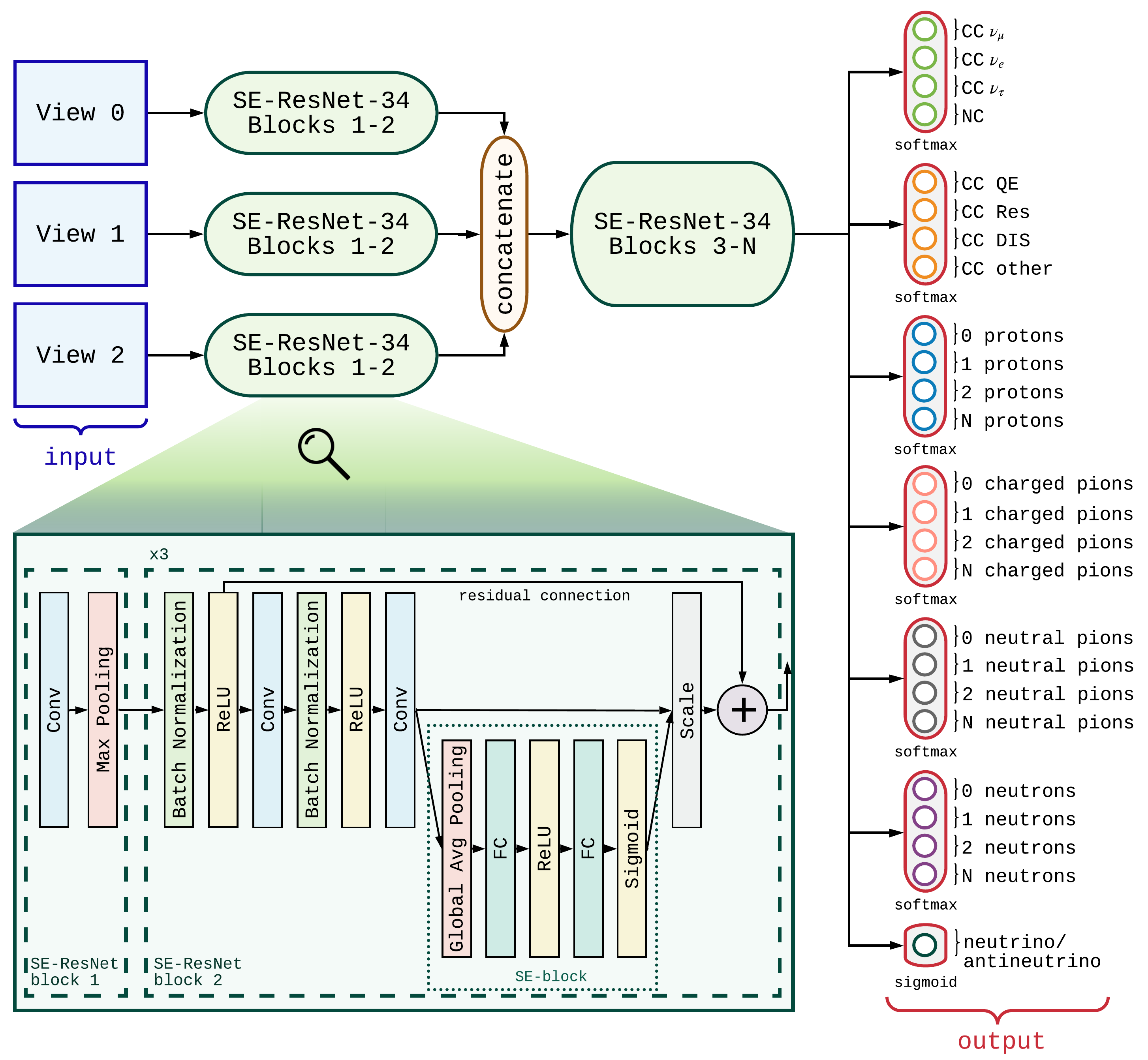}
	\caption{Simplified diagram of the DUNE CVN architecture.}
    \label{fig:cvnarchitecture}
\end{figure}

A simple overview of the architecture is shown in Fig.~\ref{fig:cvnarchitecture}. The detailed architecture of the CVN is based on the 34-layer version of the SE-ResNet architecture, which consists of a standard ResNet (Residual neural network) architecture~\cite{He-et-al-2015-deep,He-et-al-2016-identity} along with Squeeze-and-Excitation blocks~\cite{Hu-et-al-2017-squeeze}. Residual neural networks allow the $n^\textrm{th}$ layer access to the output of both the $(n-1)^\textrm{th}$ layer and the $(n-k)^\textrm{th}$ layer via a residual connection, where $k$ is a positive integer $\geq2$. This is an important feature for the DUNE CVN as it allows the fine-grained detail of a LArTPC encoded in the input images to be propagated further into the CVN than would be possible using a traditional CNN such as the GoogLeNet (also called Inception v1)~\cite{GoogLeNet} inspired network used by NOvA~\cite{novacvn}.

The DUNE CVN differs from the architectures of other residual networks discussed in the literature~\cite{He-et-al-2015-deep,He-et-al-2016-identity} in the following ways:

\begin{itemize}
    \item The input and the shallower layers of the CVN are forked into three branches - one for each view - to let the model learn parameters from each individual view (see section~\ref{sec:inputs} for more details). The outputs of the three branches are merged together by using a concatenation layer that works as input for the deeper layers of the model, as shown in Fig.~\ref{fig:cvnarchitecture}.
    
    \item The CVN returns scores for each event through seven individual outputs (see section~\ref{sec:outputs} and Fig.~\ref{fig:cvnarchitecture} for more details). Since the deeper layers of the CVN contain the model parameters\footnote{Model parameters: coefficients of the model learnt during the training stage, also known as weights.} that are simultaneously in charge of the classification for the different outputs of the network, some outputs might take advantage of the learning process of other outputs to improve their performance. Also, a multi-output network lets us weight the outputs in order to make the network pay more attention to some specific outputs (see section~\ref{sec:training} for more details). 
    
    \item Each of the three branches (blocks 1-2, the shallower layers of the architecture shown in Fig.~\ref{fig:cvnarchitecture}) consists of 7 convolutional layers, while the deeper layers (blocks 3-N in Fig.~\ref{fig:cvnarchitecture}) consist of 29 convolutional layers, making a total of 50 convolutional layers for the entire network.

\end{itemize}

\subsection{Outputs from the CVN}
\label{sec:outputs}

As shown on the right of Fig.~\ref{fig:cvnarchitecture}, there are seven outputs from the CVN, each consisting of a number of neurons with values $v_i$ for $i=1\rightarrow n$ where $n$ is the number of neurons. The sum of neuron values for each output (except for the last output since it consists of a single neuron) is given by $\sum_{i=1}^n v_i = 1$ such that each value of a neuron within a single output gives a fractional score that can be used to classify images. 

The first output, which has four neurons to classify the flavor of the neutrino interaction, is the primary output and it is the only one used in the oscillation sensitivity analysis presented in Refs~\cite{dunetdr} and~\cite{lblPaper}. The other outputs are included in the architecture for potential use in future analyses.

\begin{enumerate}
    \item The 1$^\textrm{st}$ output (4 neurons) returns scores for each event to be one of the following flavors: CC $\nu_\mu$, CC $\nu_e$, CC $\nu_\tau$ and NC. This is the primary output of the network used for the main goal of neutrino interaction flavor classification.
    \item The 2$^\textrm{nd}$ output\footnote{\label{note2}Outputs to `sub-classify' CC events. NC events are not considered for the training of this output.} (4 neurons) returns scores for each event to be one of the following interaction types: CC quasi-elastic (CC QE), CC resonant (CC Res), CC deep inelastic (CC DIS) and CC other.
    \item The 3$^\textrm{rd}$ output (4 neurons), returns scores for each event to contain the following number of protons: 0, 1, 2, $>$2.
    \item The 4$^\textrm{th}$ output (4 neurons), returns scores for each event to contain the following number of charged pions: 0, 1, 2, $>$2.
    \item The 5$^\textrm{th}$ output (4 neurons), returns scores for each event to contain the following number of neutral pions: 0, 1, 2, $>$2.
    \item The 6$^\textrm{th}$ output (4 neurons), returns scores for each event to contain the following number of neutrons: 0, 1, 2, $>$2.
    \item The 7$^\textrm{th}$ output\footnotemark[\value{footnote}] (1 neuron) returns the score for each event to be a neutrino as opposed to an antineutrino. 

\end{enumerate}

Outputs 2, 6 and 7 are not considered in the analyses presented here and are hence not further discussed, but they are included in the training and the overall loss calculations. The prediction of an event as a given underlying (anti)neutrino interaction is highly model-dependent and not as important as the number of final-state particles that can be observed in the detector, hence output 2 is not used. The neutron counting is very difficult since it is hard to define whether a neutron interaction would be visible and identifiable in the detector, so this output will not be used until it has been shown to work reliably. Finally, the antineutrino vs neutrino output is not likely to provide highly efficient or pure event selections since there is only a weak dependence on the event observables to try to differentiate neutrinos and antineutrinos.

\subsection{Training the CVN}
\label{sec:training}

The CVN\footnote{A small data release of the code is available at \url{https://github.com/DUNE/dune-cvn}.} was trained using Python 3.5.2 and Keras 2.2.4~\cite{Chollet-et-al-2015-keras} on top of Tensorflow 1.12.0~\cite{Abadi-et-al-2016-tensorflow}, on eight NVIDIA Tesla V100 GPUs. Stochastic Gradient Descent (SGD) is used as the optimizer, with a mini-batch size of 64 events (192 views), a learning rate of 0.1 (divided by 10 when the error plateaus, as suggested in~\cite{He-et-al-2015-deep}), a weight decay of 0.0001, and a momentum of 0.9\footnote{See Ref.~\cite{Goodfellow-et-al-2016-deep} for a description of optimizers and associated terminology.}. The network was trained/validated/tested on 3,212,351 events (9,637,053 images/views), consisting of 27\% CC $\nu_\mu$, 27\% CC $\nu_e$, 6\% CC $\nu_\tau$ and 40\% NC, from a single Monte Carlo sample as follows: training ($\sim98\%$), validation ($\sim1\%$) and test ($\sim1\%$). The sample of events is an MC prediction for the DUNE unoscillated FD neutrino event rate (flux times cross section) distribution in FHC beam mode as described in Ref.~\cite{dunetdr}. Samples where the input fluxes to the MC are “fully oscillated” (i.e. all $\nu_\mu$ are replaced with $\nu_e$, or all $\nu_\mu$ are replaced with $\nu_\tau$) are also used (these samples are usually weighted by oscillation probabilities and combined to produce oscillated FD event rate predictions). Analogous versions of each input sample are used for the RHC beam mode. For training purposes all CC $\nu_e$ events were considered signal since the intrinsic beam $\nu_e$ are indistinguishable from signal (appearance) $\nu_e$ at any given energy. The results presented in the following sections use a statistically independent Monte Carlo sample.

The individual loss functions for the different outputs that were used for training the model, as well as the overall loss function, are given below\footnote{Generally, $a_k$ represents the $k^\textrm{th}$ element of some vector $\bm{a}$.}:

\begin{itemize}
    \item Neutrino flavor ID, interaction type\footnote{\label{note4}A mask is applied to only consider CC events during the loss computation.}, proton count, charged pion count, neutral pion count, neutron count loss functions ($J_{1}$, $J_{2}$, $J_{3}$, $J_{4}$, $J_{5}$, and $J_{6}$, respectively): categorical cross-entropy, the loss function needed for multi-class classification.

    \begin{equation}
    J_{1} = J_{2} = J_{3} = J_{4} = J_{5} = J_{6} = - \frac{1} m \sum_{i=1}^{m} \sum_{j=1}^{c} y_{j}^{(i)} \log \hat{y}_{j}^{(i)}
    \end{equation}
    
    \item Neutrino/antineutrino ID loss function\footnotemark[\value{footnote}] ($J_7$): binary cross-entropy, the loss function needed for binary classification.

    \begin{equation}
    J_{7} = - \frac{1} m \sum_{i=1}^{m} \bm{y}^{(i)} \log (\bm{\hat{y}}^{(i)}) + (1 - \bm{y}^{(i)}) \log (1 -\bm{\hat{y}}^{(i)})
    \end{equation}

    \item Overall loss function:

    \begin{equation}
    J = \sum_{i=1}^{o} w_{i} J_{i} = w_{1} J_{1} + w_{2} J_{2} + w_{3} J_{3} + w_{4} J_{4} + w_{5} J_{5} + w_{6} J_{6} + w_{7} J_{7}
    \end{equation}

    \item Where:
        \begin{itemize}
        \item $\bm{y}^{(k)}$: true values of a specific output corresponding to the \textit{k-th} training example.
        \item $\bm{\hat{y}}^{(k)}$: predicted values of a specific output corresponding to the \textit{k-th} training example.
        \item $m$: number of training examples \{$\tX^{(1)}$, $\bm{y}^{(1)}$\}, \{$\tX^{(2)}$, $\bm{y}^{(2)}$\}, ..., \{$\tX^{(m)}$, $\bm{y}^{(m)}$\}, where $\tX^{(k)}$ means the input readout views corresponding to the $k^\textrm{th}$ training example.
        \item $c$: number of classes/neurons corresponding to a specific output $\bm{y}_{1}$, $\bm{y}_{2}$, ... , $\bm{y}_{c}$.
        \item $o$: number of outputs of the network; the CVN has seven different outputs.
        \item $\bm{w}$: output weights; vector of length $o$.
    \end{itemize}
\end{itemize}

\begin{figure}[hb]
    \centering
    \begin{tabular}{ccc}
        \begin{subfigure}{0.4\textwidth}
        \includegraphics[width=1.0\linewidth]{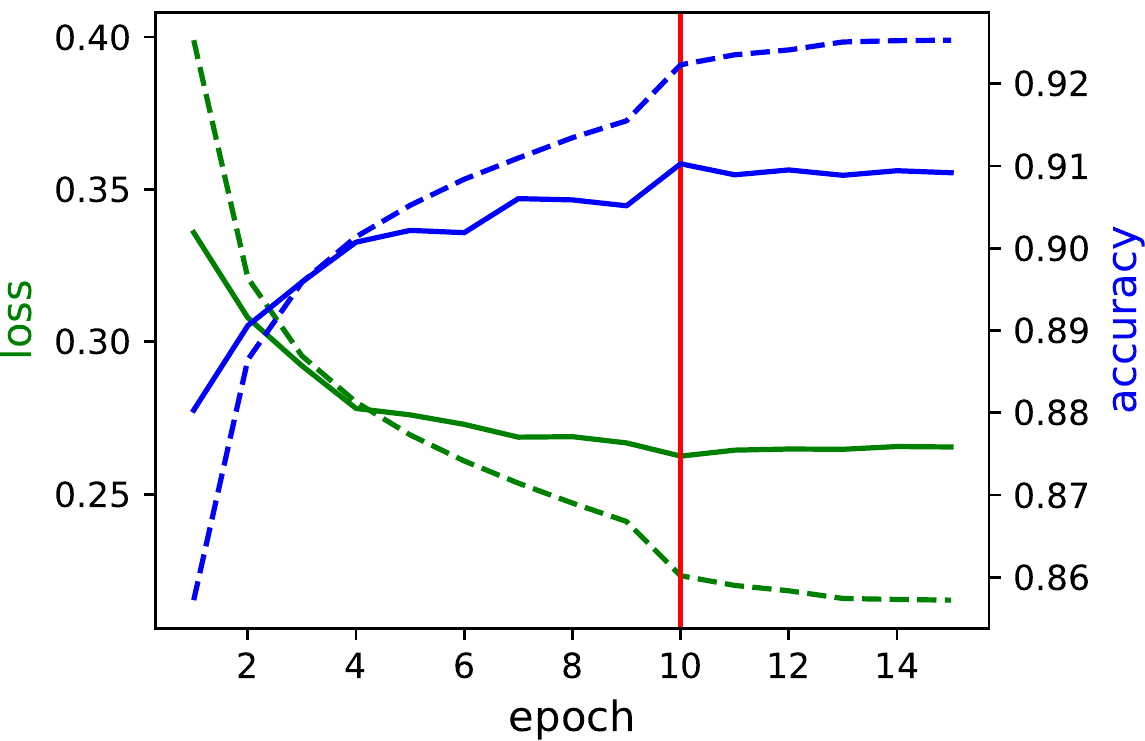}
        \caption{Flavor.} 
		\label{fig:flav_perf}
        \end{subfigure}
        \hspace*{\fill}
        \begin{subfigure}{0.4\textwidth}
        \includegraphics[width=1.0\linewidth]{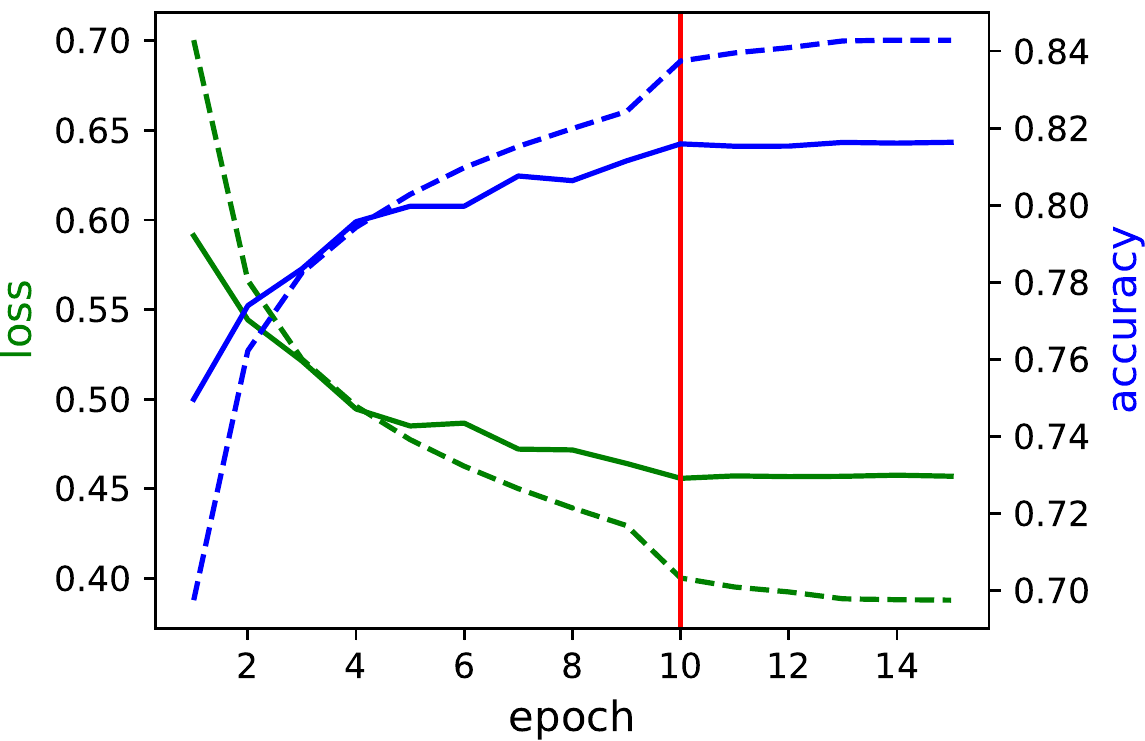}
        \caption{Protons.} 
		\label{fig:prot_perf}
        \end{subfigure}
        \\
        \begin{subfigure}{0.4\textwidth}
        \includegraphics[width=1.0\linewidth]{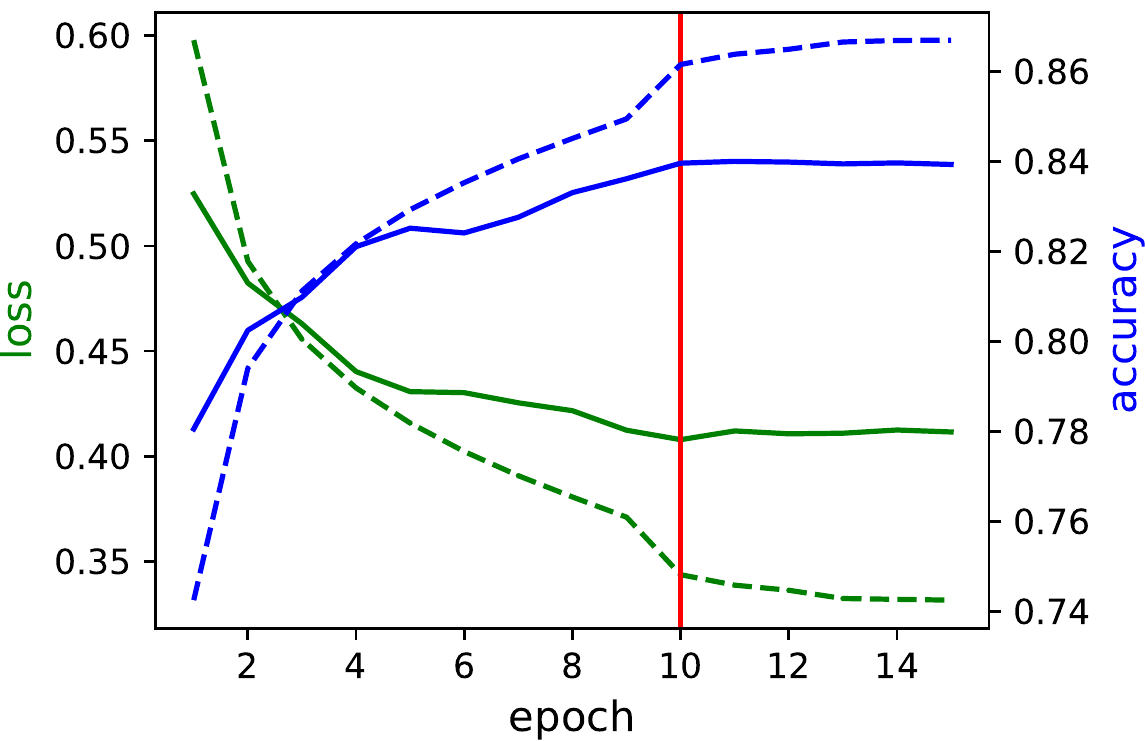}
        \caption{Charged pions.} 
		\label{fig:pion_perf}
        \end{subfigure}
        \hspace*{\fill}
        \begin{subfigure}{0.4\textwidth}
        \includegraphics[width=1.0\linewidth]{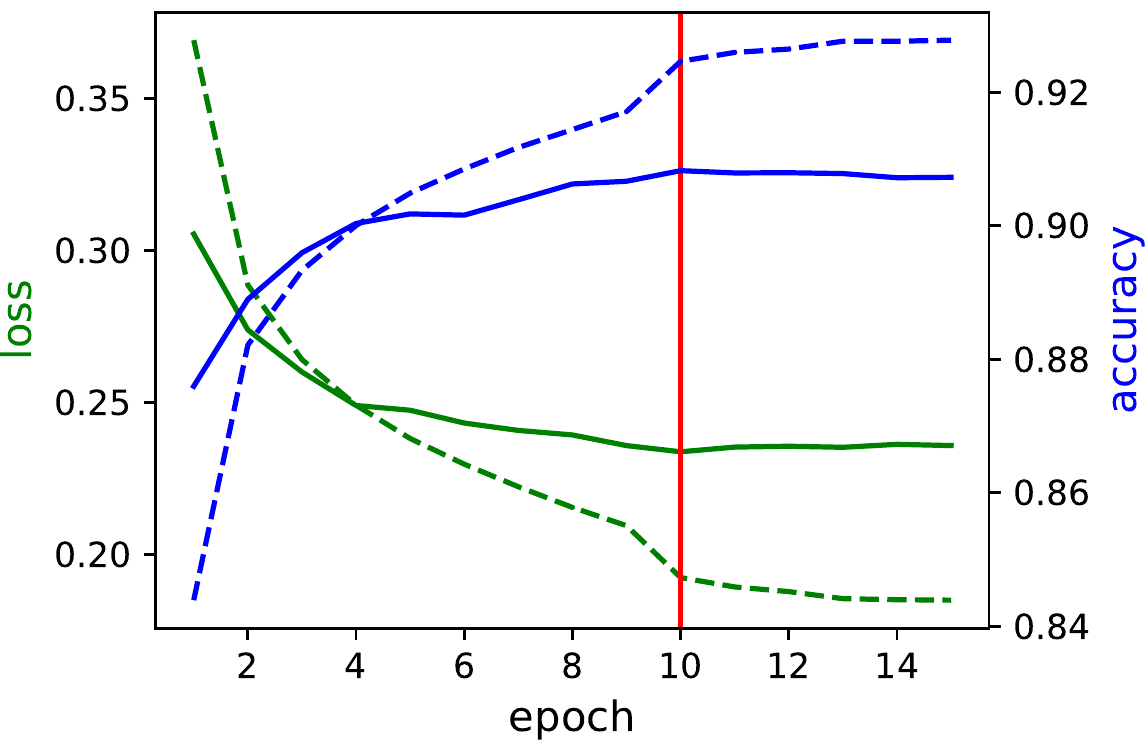}
        \caption{Neutral pions.} 
		\label{fig:pize_perf}
        \end{subfigure}
    \end{tabular}
    \caption{Loss and accuracy results for training (dashed lines) and validation (solid lines), given for the four main CVN outputs. The red vertical lines guide the eye to the network results at epoch 10, after which flavor classification performance of the validation sample does not improve.}
    \label{fig:training}
\end{figure}

The CVN was trained for 15 epochs\footnote{Epoch: one forward pass and one backward pass of all the training examples. In other words, an epoch is one pass over the entire dataset.} for $\sim$4.5 days (7 hours per epoch), and similar classification performance was obtained for the training and test samples. Figure~\ref{fig:training} shows the loss and accuracy training and validation results for the four main CVN outputs, where accuracy is defined as the fraction of events correctly classified for a given output. The red vertical lines show the epoch at which the CVN weights were taken for the model used in the presented analysis. After that epoch, the validation accuracy remains constant and small signs of overtraining begin to emerge (a small divergence of the training and validation accuracy curves). The relatively small difference between training and validation seen at epoch 10 has a negligible effect.

\subsection{Feature maps}
\begin{figure}
    \centering
		\includegraphics[width=0.94\linewidth]{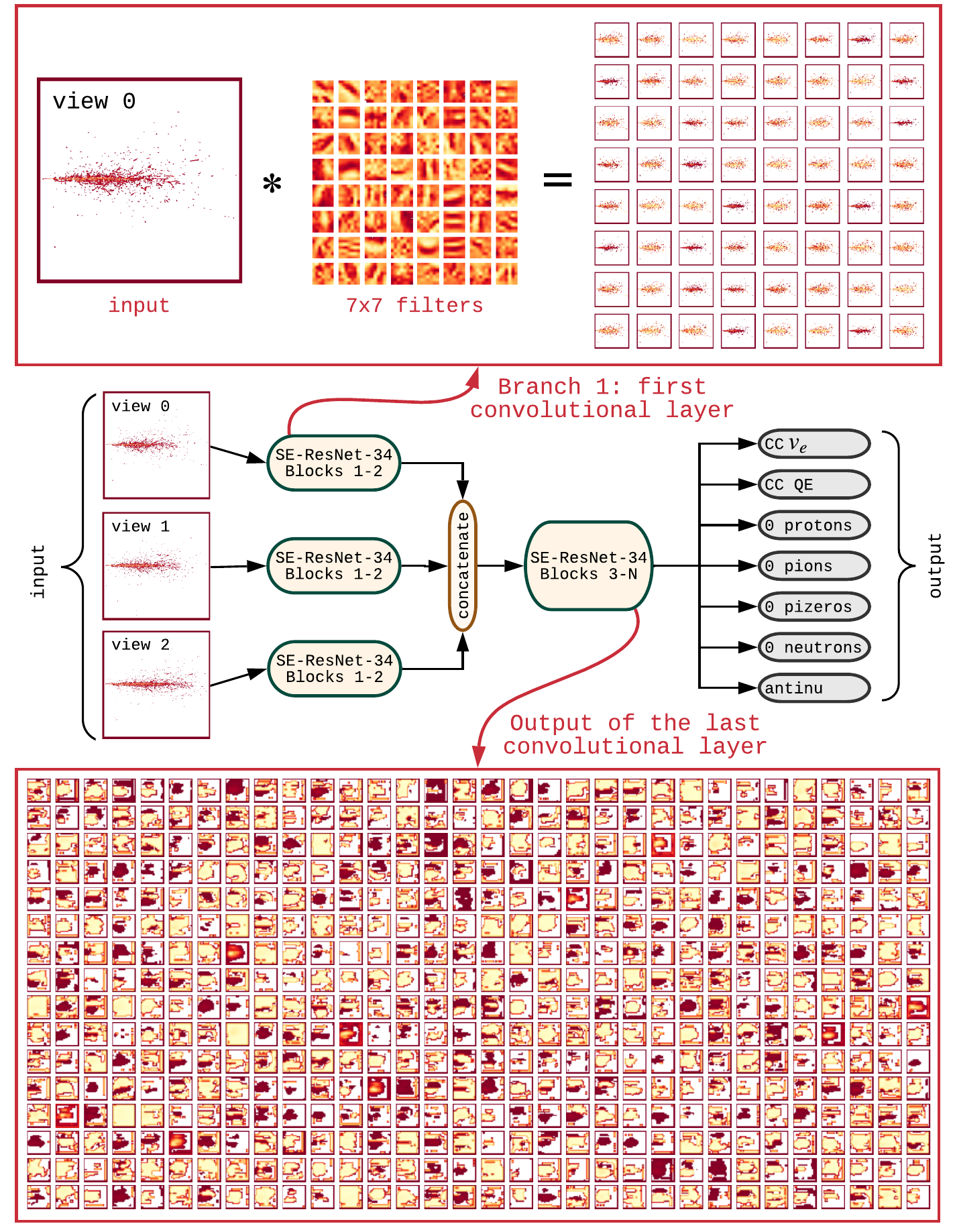}
    \caption{Visualisations of the feature extraction in the CVN for a 12.2 GeV CC $\bar{\nu}_e$ interaction. The top box shows the output from the first convolutional layer of the first branch of the network: 64 convolution kernels of size 7x7 each are applied to the image, resulting in 64 different feature maps. The bottom box shows the 512 feature maps produced by the final convolutional layer.}
    \label{fig:cvnfeaturemaps}
\end{figure}

To study how the CVN is classifying the interactions it is advisable to look at feature maps at different points in the network architecture. An example is shown in Fig.~\ref{fig:cvnfeaturemaps} for a CC $\bar{\nu}_e$ interaction, demonstrating the position from which two sets of feature maps are viewed within the network. The set of images in the top right shows the response of the filters in the first convolutional layer to the input electromagnetic shower image, where red shows a high response to a given filter, and yellow shows a low response. Across the different particle types and event topologies, the filters respond to different components in the images. The 512 feature maps from the final convolutional layer, shown at the bottom of Fig.~\ref{fig:cvnfeaturemaps} for the aforementioned CC $\bar{\nu}_e$ interaction, are much more abstract in appearance since the input images have passed through many convolutions and have hence effectively been down-sampled to a size of 16$\times$16 pixels from their original 500$\times$500 pixel size.

\section{Neutrino flavor identification performance} \label{sec:flavor}
\begin{figure}
    \centering
    \begin{tabular}{cc}
		\includegraphics[width=0.45\linewidth]{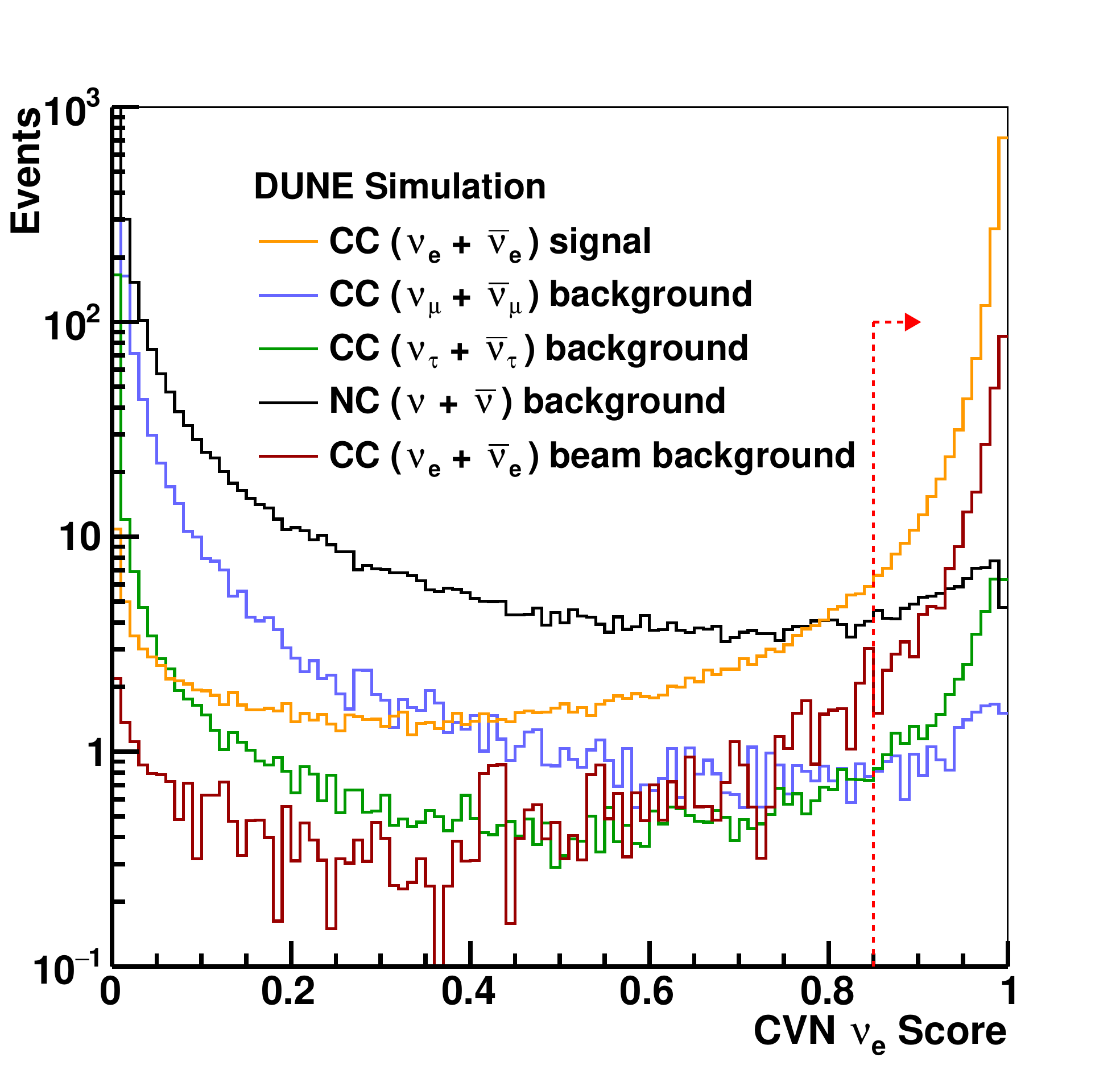} &
		\includegraphics[width=0.45\linewidth]{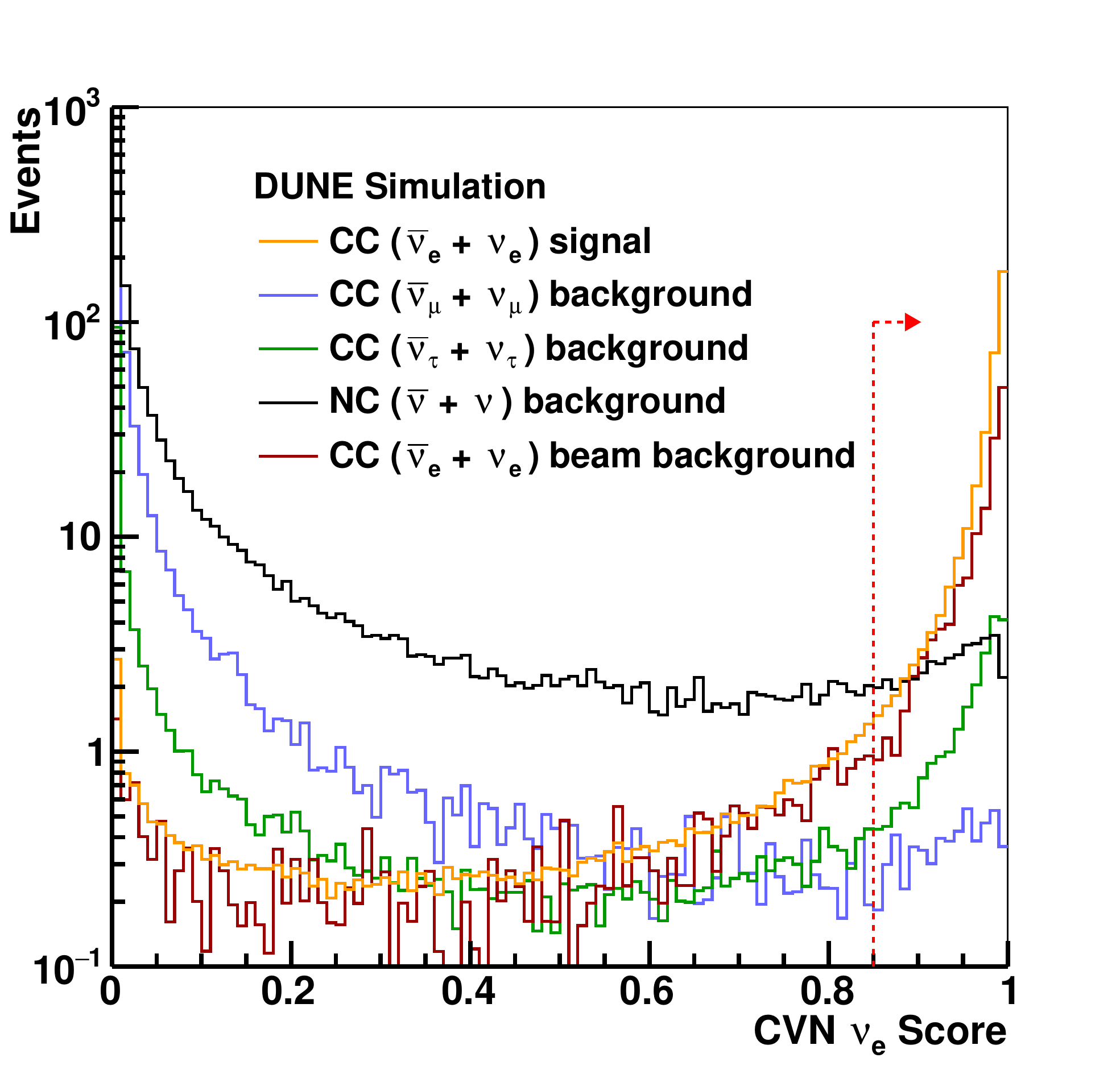} 
	\end{tabular}
	\caption{The number of events as a function of the CVN CC $\nu_e$ classification score shown for FHC (left) and RHC (right) beam modes. For simplicity, neutrino and antineutrino interactions have been combined within each histogram category. A log scale is used on the y-axis, normalized to 3.5 years of staged running, and the arrows denote the cut values applied for the DUNE TDR analyses~\cite{dunetdr}.}
    \label{fig:cvnprobnue}
\end{figure}

\begin{figure}
    \centering
    \begin{tabular}{cc}
		\includegraphics[width=0.45\linewidth]{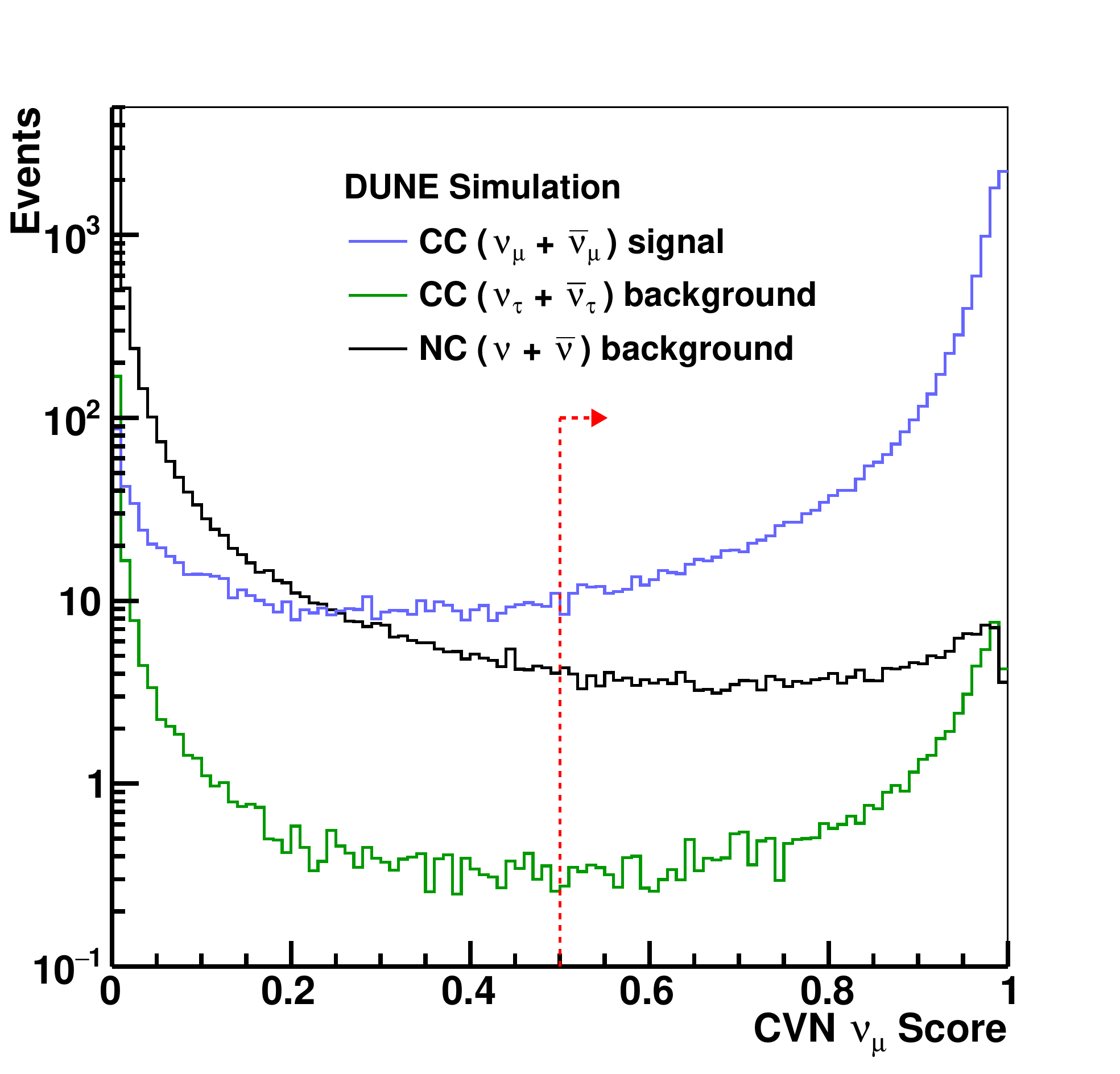} &
		\includegraphics[width=0.45\linewidth]{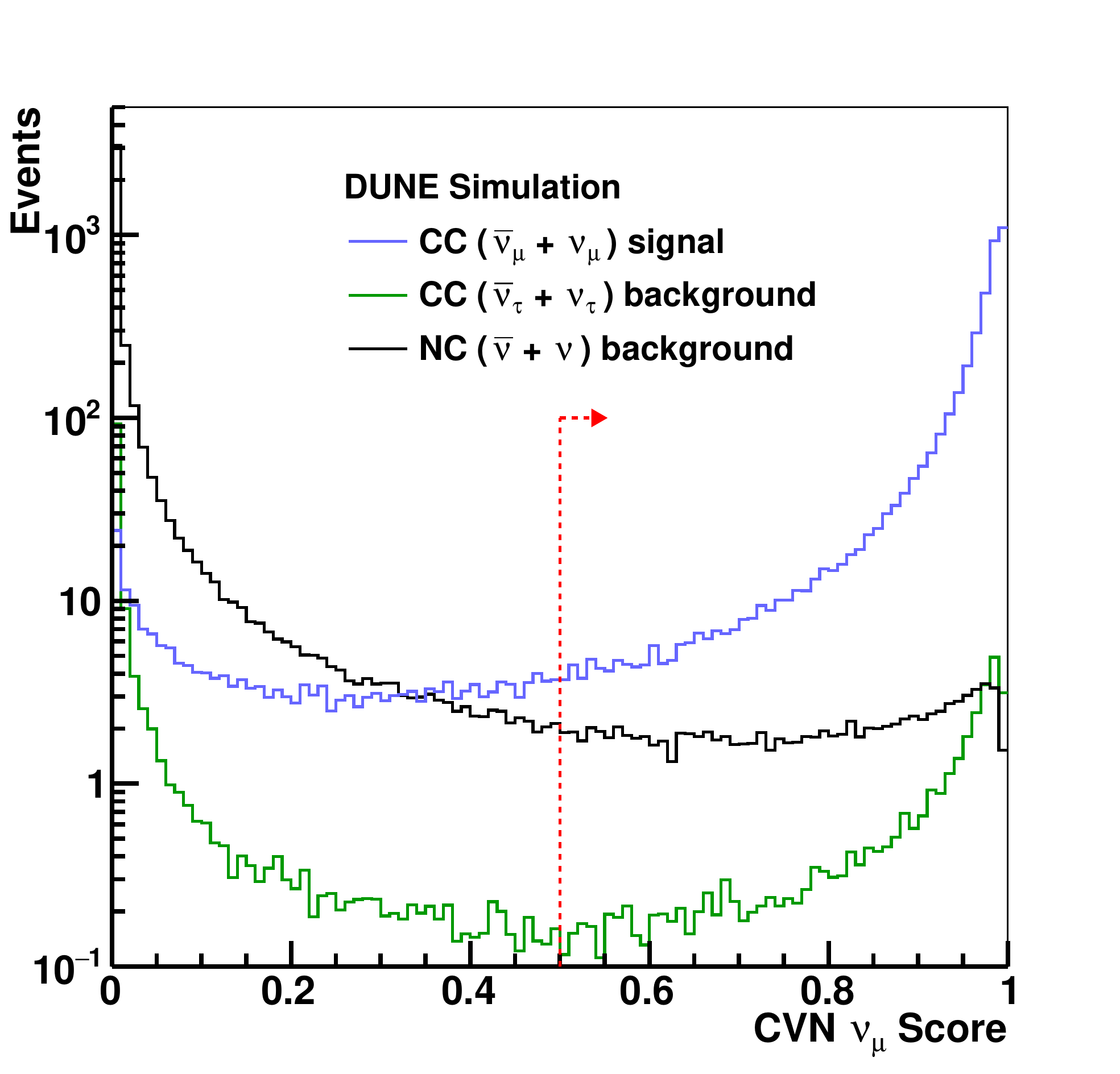} 
	\end{tabular}
    \caption{The number of events as a function of the CVN CC $\nu_\mu$ classification score shown for FHC (left) and RHC (right) beam modes. For simplicity, neutrino and antineutrino interactions have been combined within each histogram category. Backgrounds from CC $\nu_e$ interactions are negligible and not shown. A log scale is used on the y-axis, normalized to 3.5 years of staged running, and the arrows denote the cut values applied for the DUNE TDR analyses~\cite{dunetdr}.}
    \label{fig:cvnprobnumu}
\end{figure}

The primary goal of the CVN is to accurately identify CC $\nu_e$, CC $\bar{\nu}_e$, CC $\nu_\mu$ and CC $\bar{\nu}_\mu$ interactions for the selection of the samples required for the neutrino oscillation analysis. The values of the neurons in the flavor output give the score for each neutrino interaction to be one of the neutrino flavors. The CVN CC $\nu_e$ score distribution, $P(\nu_e)$, is shown for the FHC beam mode (left) and RHC (right) in Fig.~\ref{fig:cvnprobnue} for all interactions with a reconstructed event vertex within the FD fiducial volume, as described in Ref.~\cite{lblPaper}. The contributions from neutrino and antineutrino components for each flavor are combined since the detector can not easily distinguish between them. Very clear separation is seen between the signal (CC $\nu_e$ and CC $\bar{\nu}_e$) interactions and the background interactions including those from NC $\nu$ and NC $\bar{\nu}$ events. The beam CC $\nu_e$ background is seen to peak in the same way as the CC $\nu_e$ signal, which is expected since both arise from the same type of neutrino interaction. Figure~\ref{fig:cvnprobnumu} shows the corresponding plots for $P(\nu_\mu)$ for FHC and RHC beam modes for the same set of interactions. In all four histograms the signal interactions are peaked closely near score values of unity and the backgrounds lie close to zero score, as expected.

The CC $\nu_e$ event selection criteria are chosen to maximize the oscillation analysis sensitivity to $CP$-violation; i.e.: significance of the determination that $\sin\left(\delta_{CP}\right)\ne0$~\cite{lblPaper}. The optimization was performed using a simple scan of cuts on $P(\nu_e)$ for a single true value of $\delta_{CP}$. $CP$-violation sensitivity does not strongly depend on the selection criterion for $P(\nu_\mu)$ so this cut was chosen by inspection of Fig.~\ref{fig:cvnprobnumu}. The resulting requirements are $P(\nu_e) > 0.85$ for an interaction to be selected as a CC $\nu_e$ candidate and  $P(\nu_\mu) > 0.5$ for an interaction to be selected as a CC $\nu_\mu$ candidate. These cut values are represented by the red arrows in Figs. 7 and 8. Since all of the flavor classification scores must sum to one, these two samples are mutually exclusive. The same CVN and selection criteria are used for both FHC and RHC event selections.

\begin{figure}
    \centering
		\includegraphics[scale=0.3]{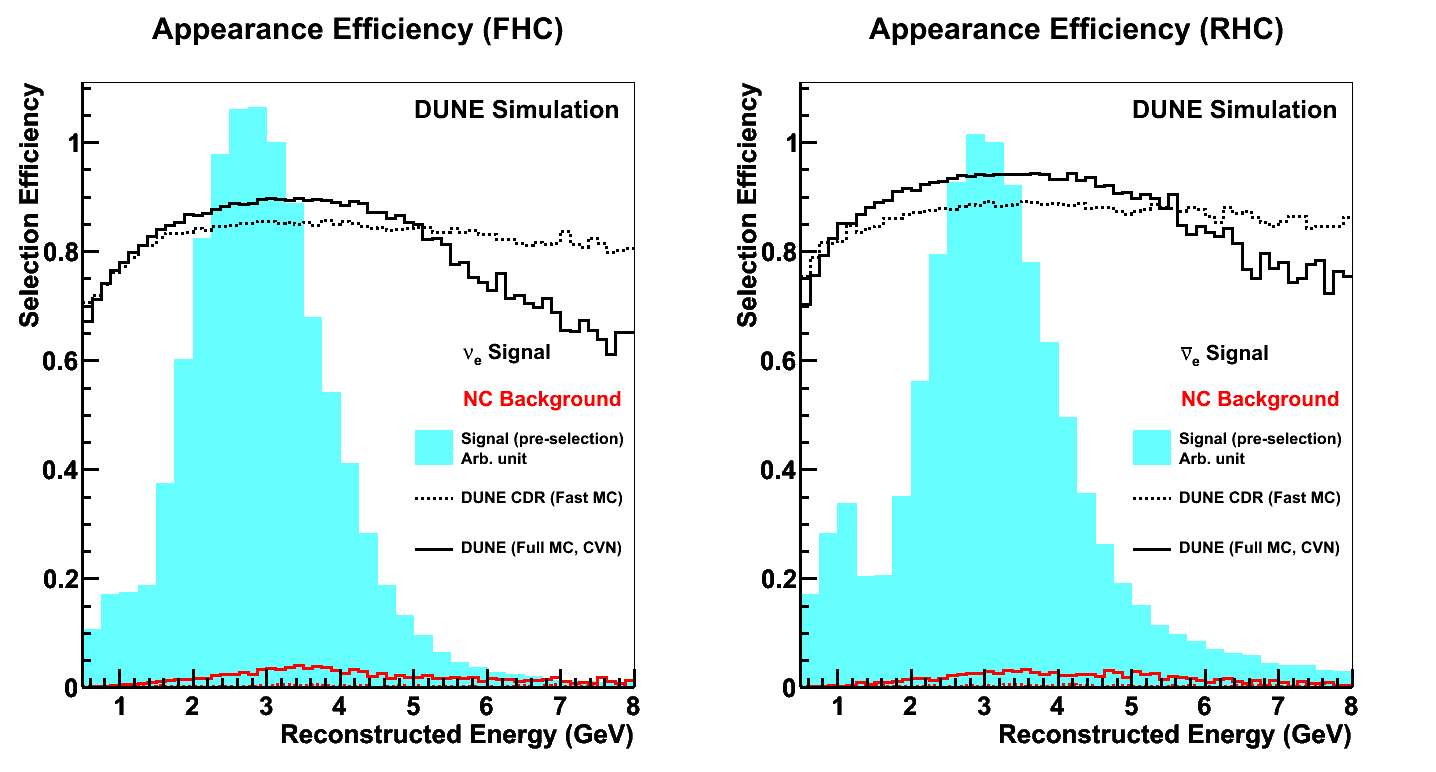} 
	\caption{The CC $\nu_e$ selection efficiency for FHC-mode (left) and RHC-mode (right) simulation with the criterion $P(\nu_e) > 0.85$. The solid (dashed) lines show results from the CVN (CDR) for signal CC $\nu_e$ and CC $\bar{\nu}_e$ events in black and NC background interactions in red. The cyan shaded region shows the oscillated flux to illustrate the most important regions of the energy distribution.}
    \label{fig:nueeff}
\end{figure}

\begin{figure}
    \centering
		\includegraphics[scale=0.3]{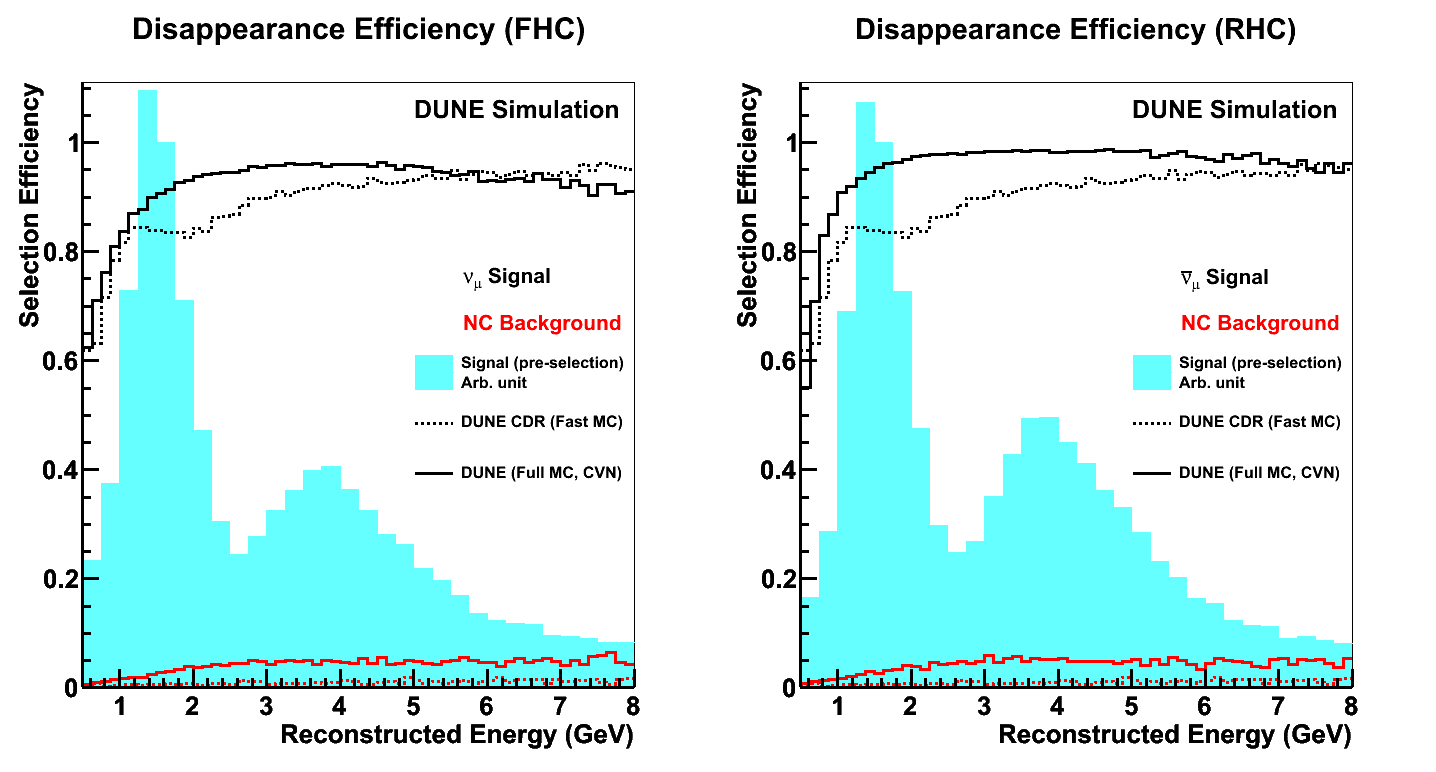} 
	\caption{The CC $\nu_\mu$ selection efficiency for FHC-mode (left) and RHC-mode (right) simulation with the criterion $P(\nu_\mu) > 0.5$. The solid (dashed) lines show results from the CVN (CDR) for signal CC $\nu_\mu$ and CC $\bar{\nu}_\mu$ events in black and NC background interactions in red. The cyan shaded region shows the oscillated flux to illustrate the most important regions of the energy distribution.}
    \label{fig:numueff}
\end{figure}

Figure~\ref{fig:nueeff} shows the efficiency as a function of reconstructed energy (under the electron neutrino hypothesis, as discussed in Section~\ref{sec:reco}) for the CC $\nu_e$ and CC $\bar{\nu}_e$ event selections. The efficiency for the CVN is shown compared to the predicted efficiency used in the DUNE Conceptual Design Report (CDR)~\cite{duneCDR}, demonstrating that, across the most important part of the flux distribution (less than 5\,GeV), the performance can exceed the CDR assumption. The efficiency in FHC (RHC) mode peaks at 90\% (94\%) and exceeds 85\% (90\%) for reconstructed neutrino energies between 2-5\,GeV. Antineutrino interactions, on average, produce more energetic leptons and fewer hadrons than neutrino events, leading to greater lepton tagging efficiency with respect to neutrino-induced events. The training was optimized over the oscillation peak between 1\,GeV and 5\,GeV, and hence the CVN performs best in this region where the sensitivity to neutrino oscillations is greatest. Improvements to the efficiency above 5 GeV may be achieved through the inclusion of more relevant training data, but requires more study. The CDR analysis was based on a fast simulation that employed a parameterized detector response based on GEANT4 single particle simulations, and a classification scheme that classified events based on the longest muon/charged pion track, or the largest EM shower if no qualifying track was present. The efficiencies at low energy were tuned to hand scan results as a function of lepton energy and event inelasticity. Figure~\ref{fig:numueff} shows the corresponding selection efficiency for the CC $\nu_\mu$ event selection. The efficiency has a maximum efficiency of 96\% (97\%) and exceeds 90\% (95\%) efficiency for reconstructed neutrino energies above 2\,GeV for the FHC (RHC) beam mode. The optimized cut values permit a larger background component than the CDR analysis but the overall performance of the selection is increased due to the significantly improved signal efficiency. Considering all electron neutrino interactions (both appeared and beam background CC $\nu_e$ and CC $\bar{\nu}_e$ events) as signal interactions, the CVN has a selection purity of 91\% (89\%) for the FHC (RHC) beam mode, assuming the normal neutrino mass ordering and $\delta_{CP} = 0$~\cite{lblPaper}. 

\section{Exclusive final state results} \label{sec:count}
The CVN has three outputs that count the number of final-state particles for the following species: protons, charged pions, and neutral pions. Neutrino interactions with different final-state particles can have different energy resolutions and systematic uncertainties depending on the complexity and particle multiplicity of the interaction. It may be possible to improve the oscillation sensitivity of the analysis by identifying subsamples of events with specific interaction topologies and very good energy resolution.

\begin{figure}
    \centering
    \begin{tabular}{cc}
		\includegraphics[width=0.475\linewidth]{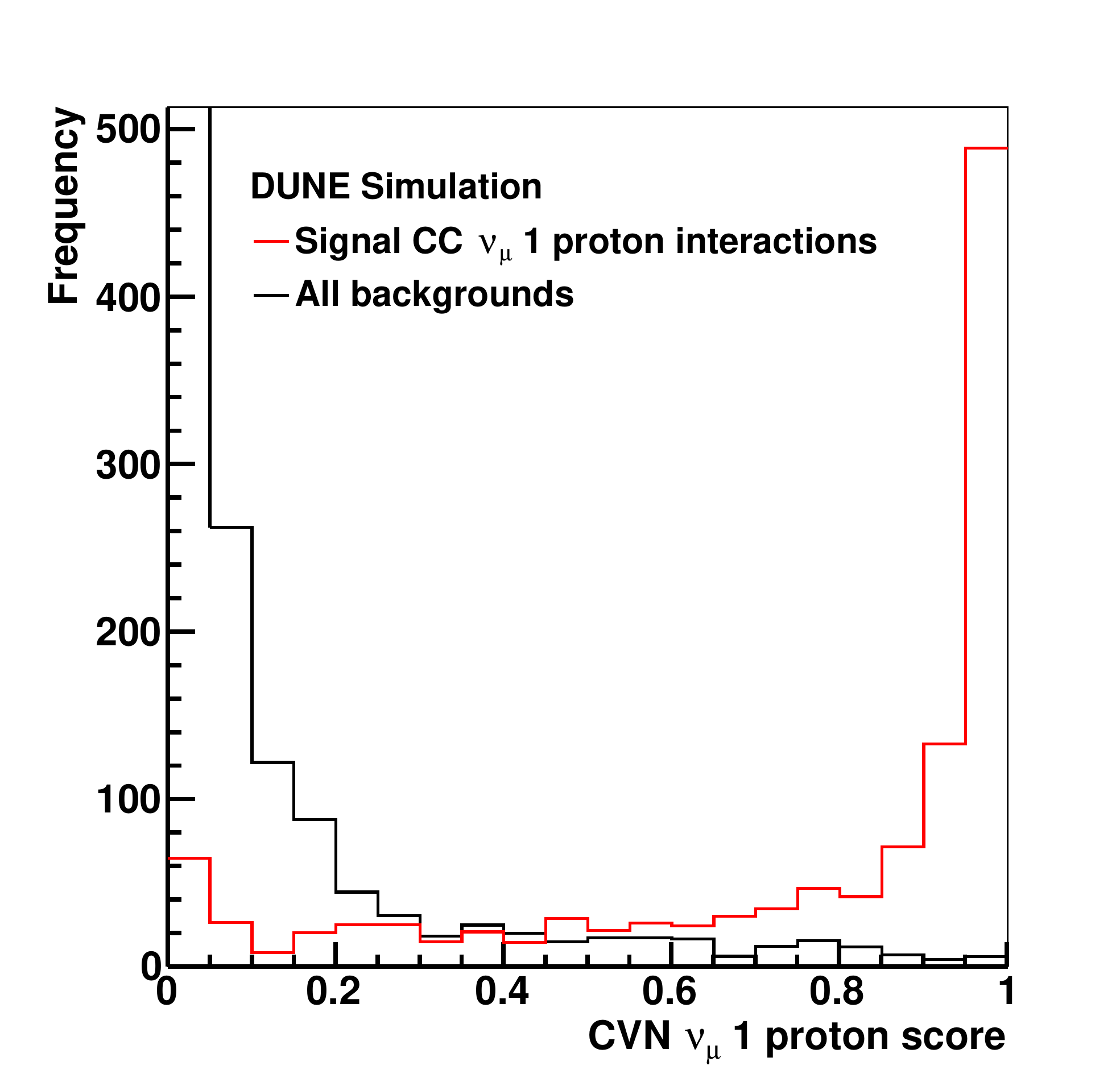} &
		\includegraphics[width=0.475\linewidth]{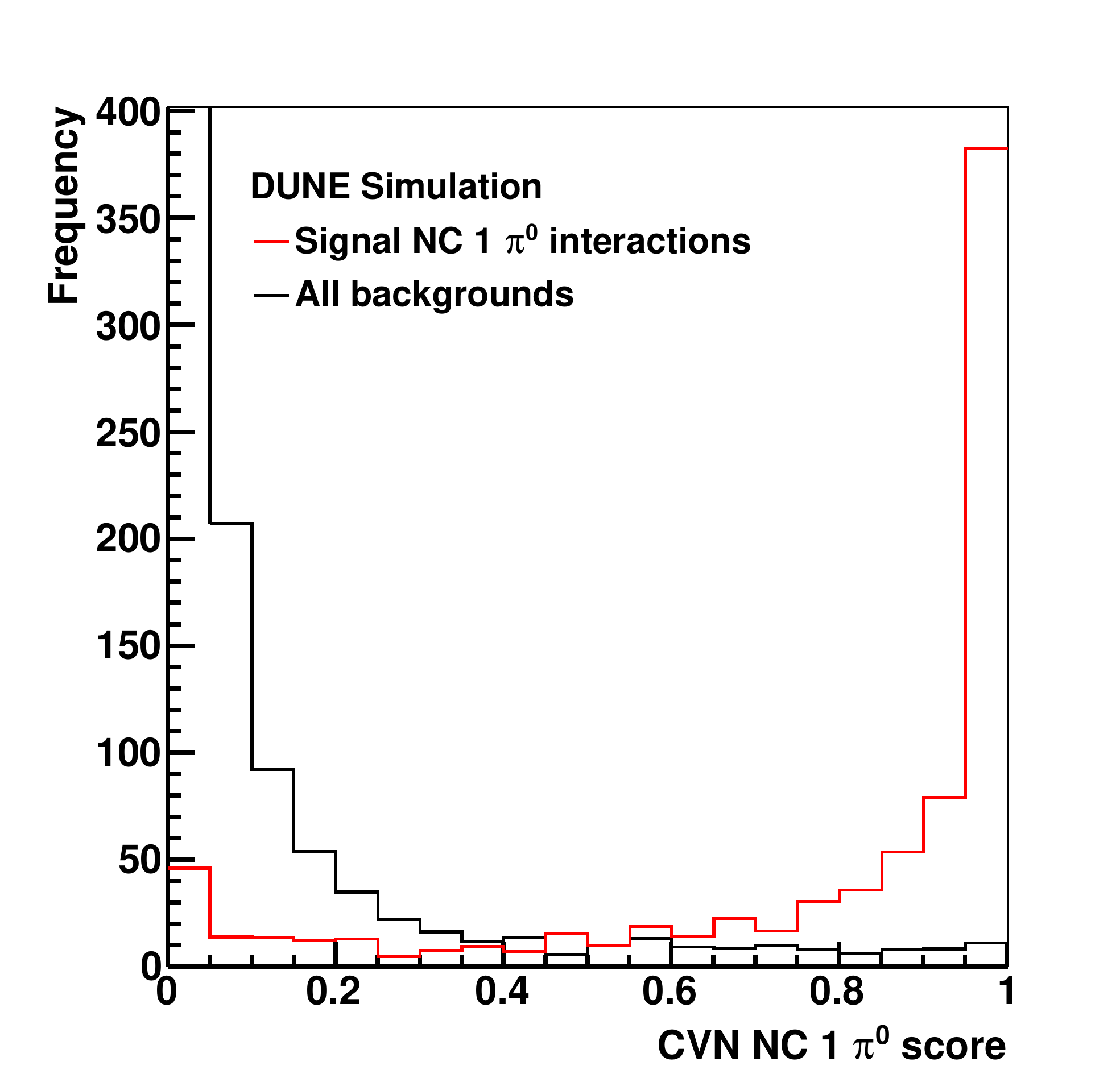}
	\end{tabular}
	\caption{The CC $\nu_\mu$ 1 proton (left) and NC 1$\pi^0$ (right) combined score distributions from the CVN. In both cases the number of other particles is required to be zero. All events that do not fit the signal description comprise the all backgrounds histograms. The histograms are shown in the expected relative fractions but the overall scale is arbitrary.}
    \label{fig:exclusive}
\end{figure}

The individual output scores from the CVN can be multiplied together to give compound scores for exclusive selections. For example, the left plot in Fig.~\ref{fig:exclusive} shows the combined score for an event to be CC $\nu_\mu$ with only a single proton in the final-state hadronic system, formed by the product 
\begin{equation}
P\left(\textrm{CC}\nu_\mu\,1\textrm{p}\right) = P\left(\nu_\mu \right) P\left( 1\,\textrm{p} \right) P\left( 0 \pi^\pm \right)P\left( 0 \pi^0 \right).
\end{equation}
Similarly, the right plot of Fig.~\ref{fig:exclusive} shows NC$1\pi^0$ score, which contains only a single visible $\pi^0$ meson in the final state, defined as: 
\begin{equation}
P\left(\textrm{NC}\,1\pi^0\right) = P\left(\textrm{NC}\right) P\left(0\textrm{p} \right) P\left(0\pi^\pm \right) P\left(1\pi^0 \right).
\end{equation}
The background and signal distributions, closely peaked toward 0 and 1 respectively, demonstrate that the efficient selection of exclusive final states will be possible with the DUNE CVN technique. However, it is possible that the CVN is keying in on features of the model that are not well-supported by data (e.g. kinematic distributions of particles in the hadronic shower) rather than well-supported features, like the individual particle energy deposition patterns. Studies of potential bias from selections based on these classifiers are required before they can be used to generate analysis samples. Provided that the particle counting outputs can be shown to work in a robust manner
for simulations and experimental data, these detailed selections have the potential to significantly improve the scientific output of DUNE FD data.

\section{Robustness}
A common concern about the applications of deep learning in high energy physics is the difference in performance between data and simulation. A straightforward check of the CVN robustness is to inspect plots of the CVN efficiency as a function of various kinematic quantities. More advanced studies could be imagined where the underlying input physics model is changed to produce alternate input samples for training and testing purposes. Studies of this nature are beyond the scope of this paper, but should be part of the validation scheme for any deep learning discriminant used in eventual analyses of DUNE data.

To be considered well-behaved, the CVN flavor identification
should be sensitive to the presence of a visible charged lepton and not highly dependent on the details of the hadronic system, which could be poorly modeled. A visible charged lepton requires that the track or shower that it produces has clearly distinguishable features that are not masked by the presence of many overlapping energy depositions from particles from the hadronic shower. Furthermore, background interactions selected by the CVN should be those containing charged pions (for CC $\nu_\mu$) or neutral pions (for CC $\nu_e$) that mimic the charged leptons in the signal interactions. Plots of selection efficiency for signal and background interactions were generated as a function of a variety of true and reconstructed quantities, several of which are highlighted here.

\begin{figure}
    \centering
    \begin{tabular}{cc}
		\includegraphics[width=0.475\linewidth]{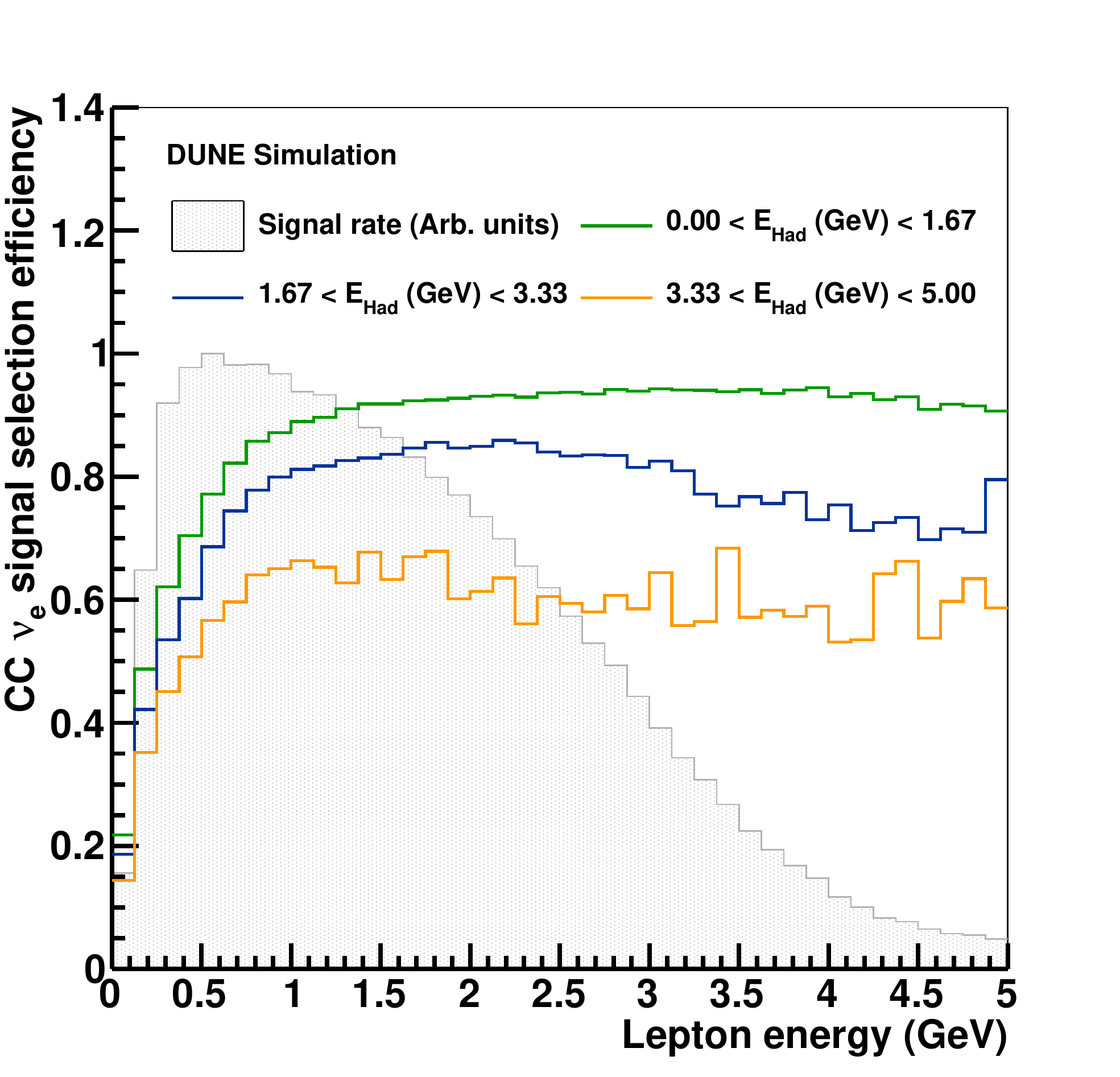} &
		\includegraphics[width=0.475\linewidth]{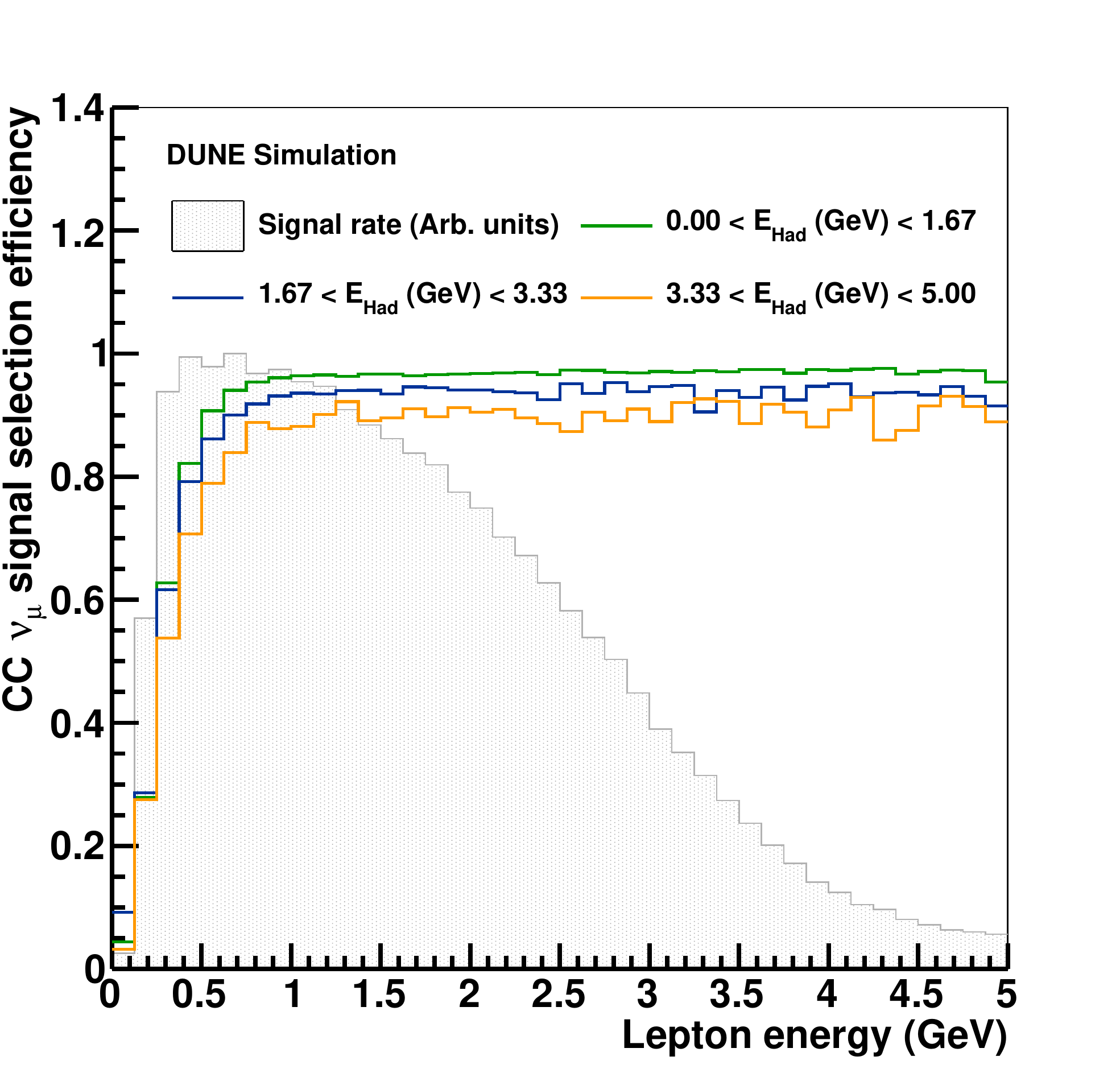}
	\end{tabular}
    \caption{The variation of the signal selection efficiency as a function of the charged lepton energy shown for three ranges of hadronic energy, $E_\textrm{Had}$, for the CC $\nu_e$ (left) and CC $\nu_\mu$ (right) selections. In both cases the distribution of the unoscillated signal events before selection is shown by the filled grey histogram.}
    \label{fig:signal_eff}
\end{figure}

Figure~\ref{fig:signal_eff} shows the variation of the signal selection efficiency as a function of the charged lepton energy for three ranges of hadronic energy for the CC $\nu_e$ (left) and CC $\nu_\mu$ (right) selections. There is a threshold around 0.1\,GeV below which no events are correctly identified, and a region at higher lepton energy where the efficiency reaches a maximum and remains relatively flat. As the hadronic energy increases the maximum efficiency decreases, and this effect is more pronounced for the CC $\nu_e$ selection since EM showers are more easily masked by hadronic shower energy depositions, as compared with long, straight muon tracks. The CC $\nu_\mu$ efficiency as a function of the true muon energy also demonstrates that the performance is not affected by the lack of confinement of higher energy muons ($\gtrsim1\,$GeV) within the $500\times500$ pixel images, as was discussed in Section~\ref{sec:inputs}.


\begin{figure}
    \centering
    \begin{tabular}{cc}
		\includegraphics[width=0.475\linewidth]{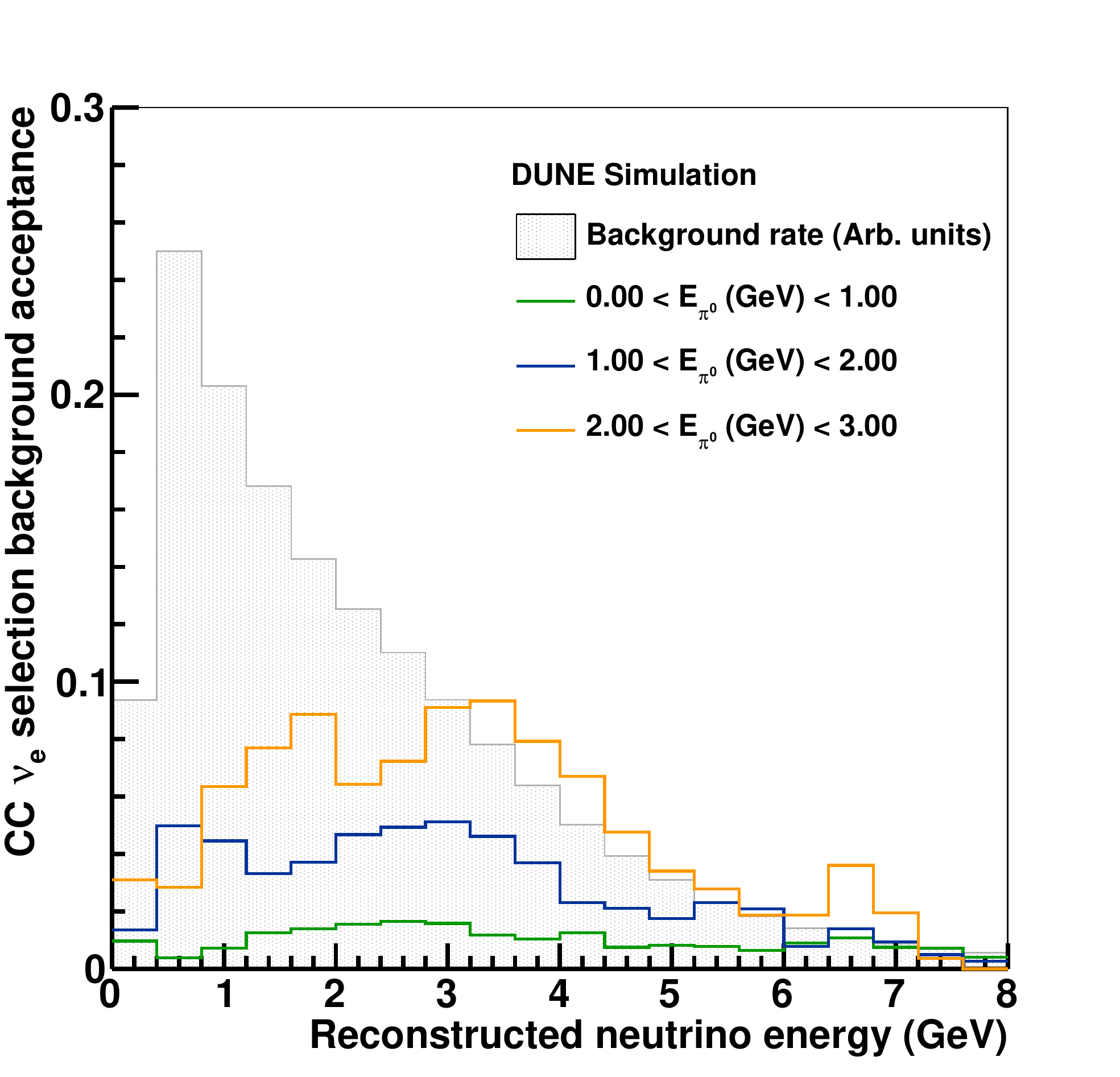} &
		\includegraphics[width=0.475\linewidth]{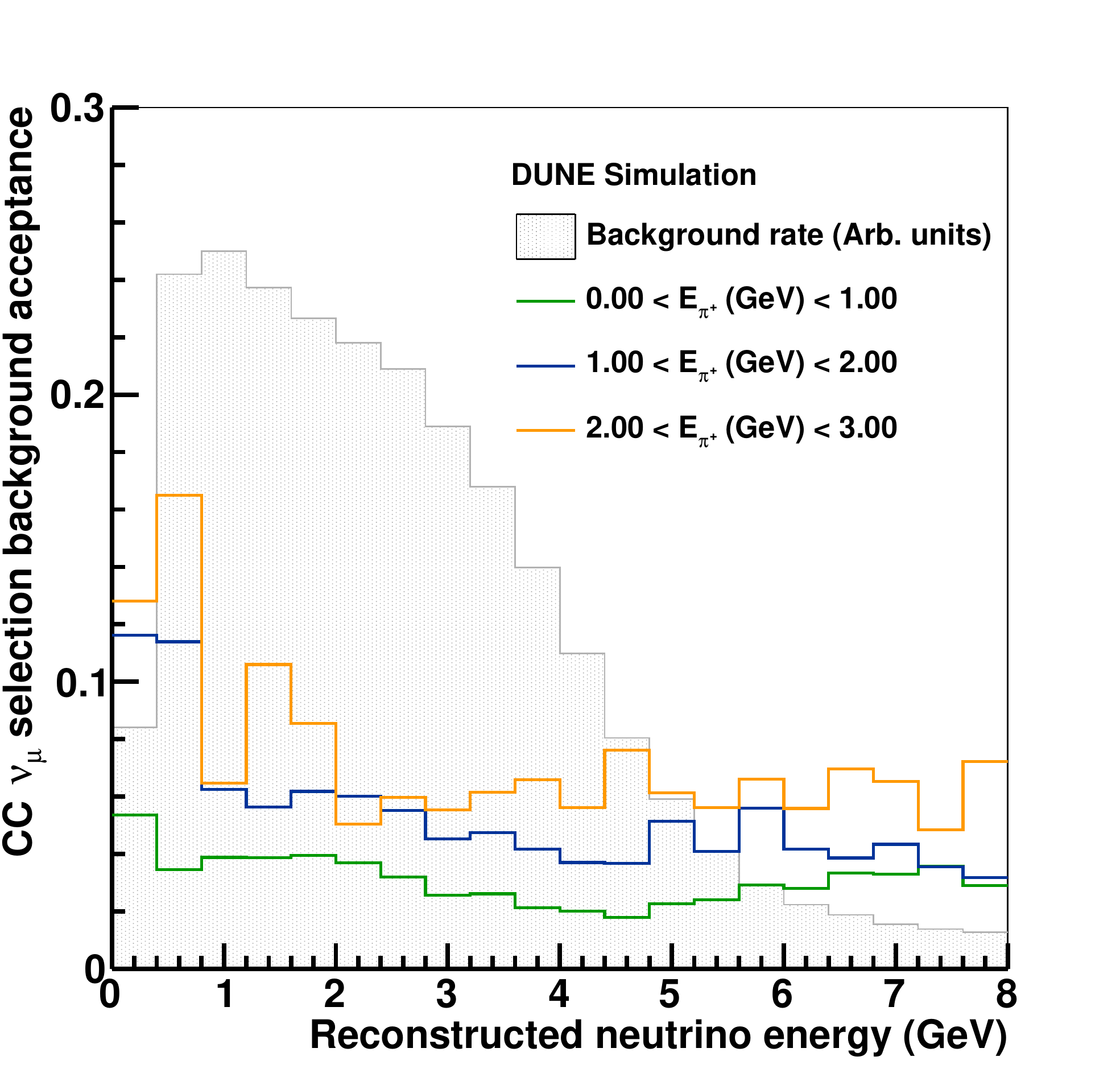}
	\end{tabular}
    \caption{Left: the background acceptance in the CC $\nu_e$ selection for background interactions containing a final-state $\pi^0$ meson. Right: the background acceptance in the CC $\nu_\mu$ selection for background interactions containing a final-state $\pi^+$ meson. In both cases the energy distribution of the unoscillated backgrounds before selection is shown by the filled grey histogram.}
    \label{fig:background_eff}
\end{figure}

The plot on the left of Fig.~\ref{fig:background_eff} shows the efficiency in the CC $\nu_e$ selection for background interactions containing a $\pi^0$ meson as a function of the reconstructed $\nu_e$ energy distribution for three ranges of $\pi^0$ energy, $E_{\pi^0}$. As expected, the selection efficiency is larger for the background interactions with higher energy $\pi^0$ mesons. Similarly, the selection efficiency for background interactions containing a $\pi^+$ meson in the CC $\nu_\mu$ selection is shown on the right of Fig.~\ref{fig:background_eff} to be larger for higher energy mesons.


\begin{figure}
    \centering
    \includegraphics[width=0.475\linewidth]{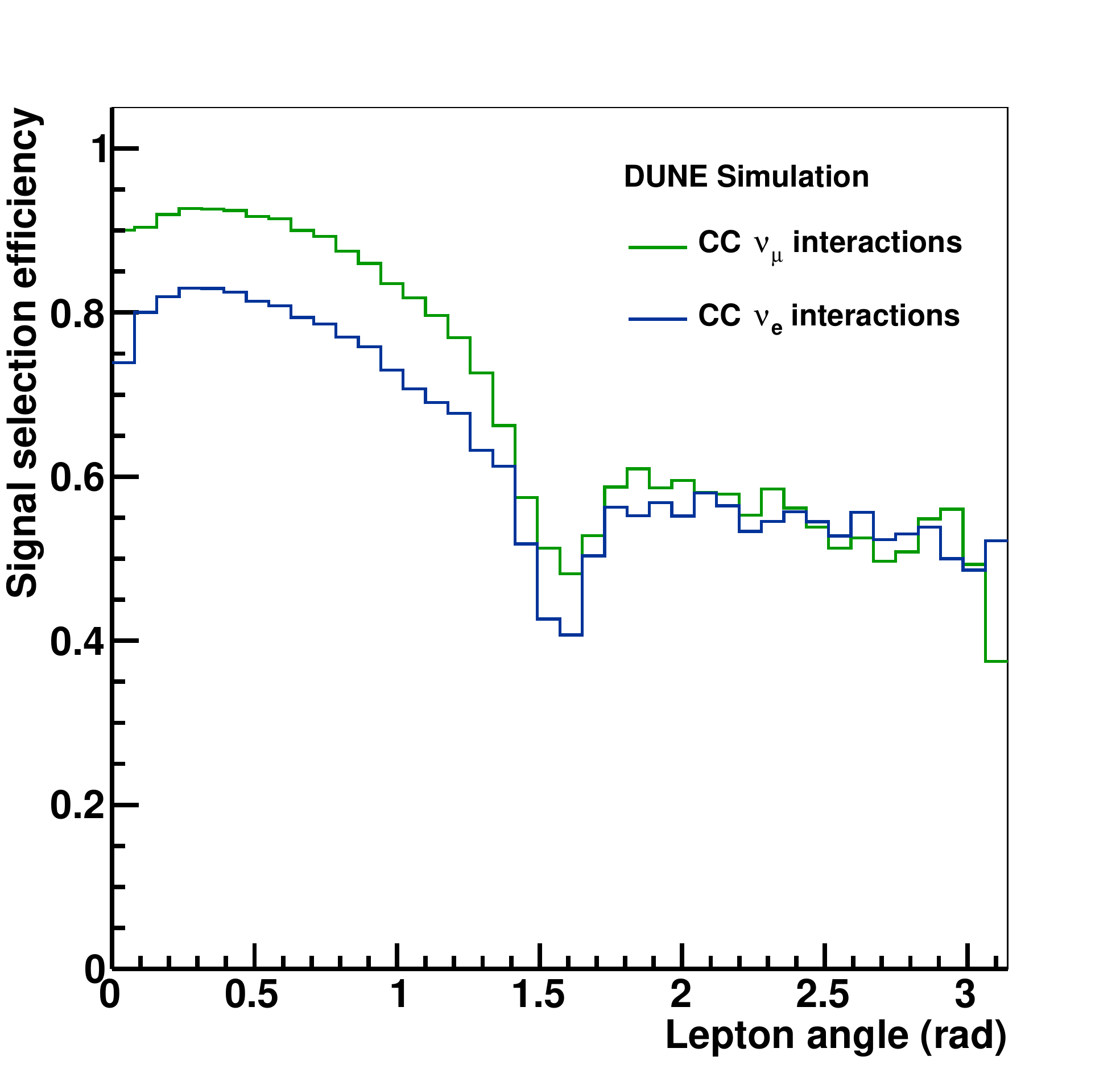}
    \caption{The signal selection efficiency for CC $\nu_\mu$ (green) and CC $\nu_e$ (blue) interactions as a function of the angle between the outgoing charged lepton and the parent neutrino.}
    \label{fig:signal_eff_angle}
\end{figure}

Figure~\ref{fig:signal_eff_angle} shows the selection efficiency for CC $\nu_\mu$ and CC $\nu_e$ interactions as a function of the charged lepton angle, defined with respect to the neutrino direction. This angle is defined in 3D, hence when the angle is $90^\circ$ it corresponds to two cases where the efficiency is expected to be lower: the lepton is travelling almost perpendicular to the readout planes, or the lepton is travelling parallel to the collection plane (view 2) wires. In these two cases the CVN does not have clear images of the charged lepton in one or more readout views. This angle is also strongly correlated with the charged lepton energy, explaining the lower efficiency for events containing backward going, and hence lower energy, charged leptons.

Additional studies, not shown here, help to elucidate other features of these distributions. For example, a small fraction of events with very low energy leptons are still correctly identified. For these events it can be shown that they contain high energy pions which are likely responsible for their strong CVN flavor identification scores. Also of note are studies of the efficiency for other kinematic variables that showed no dependence other than those induced by their correlations with the leptonic and hadronic system energies. Finally, studies of CC $\nu_\tau$ events showed that efficiencies were consistent with the tau decay rates to muons and electrons. Roughly 17\% of CC $\nu_\tau$ events were classified as CC $\nu_\mu$, and about 17\% as CC $\nu_e$. The primary $\tau^\pm$ decays before leaving a track in the detectors, and though CC $\nu_\tau$ event kinematics are different from CC $\nu_e$ and CC $\nu_\mu$ events, these events are classified based on the visible charged lepton in the event.

The outcome of these studies provides confidence that the CVN classification is strongly tied to the charged lepton features: EM showers and muon tracks. The lowest performance is seen for indistinguishable intrinsic backgrounds, such as beam-induced electron neutrinos, and events with a misidentified hadron and no visible, lepton-induced track or shower.


\section{Conclusion}
The DUNE CVN algorithm provides excellent neutrino flavor classification, reaching efficiencies of 90\% for electron neutrinos and 95\% for muon neutrinos. These efficiencies have basic features that are consistent with those presented in the DUNE CDR~\cite{duneCDR}. The CVN outperforms the CDR estimates, exceeding the signal selection efficiency over most of the energy ranges shown, albeit with slightly decreased background rejection capability. The results presented here form a key part of the neutrino oscillation analysis sensitivities presented in the DUNE TDR~\cite{dunetdr}. A proof-of-principle demonstration of final-state particle counting showed a potential mechanism by which to subdivide the event selections to further improve the analysis sensitivity. Future studies of possible systematic biases arising from physics models are planned to ensure the robustness of the particle counting outputs.

\section{Acknowledgements}
This document was prepared by the DUNE collaboration using the
resources of the Fermi National Accelerator Laboratory 
(Fermilab), a U.S. Department of Energy, Office of Science, 
HEP User Facility. Fermilab is managed by Fermi Research Alliance, 
LLC (FRA), acting under Contract No. DE-AC02-07CH11359.
%
%
This work was supported by
CNPq, FAPERJ, FAPEG and FAPESP,              Brazil;
CFI, IPP and NSERC,                          Canada;
CERN;
M\v{S}MT,	                                 Czech Republic;
ERDF, H2020-EU and MSCA,                     European Union;
CNRS/IN2P3 and CEA,                          France;
INFN,                                        Italy;
FCT,                                         Portugal;
NRF,                                         South Korea;
CAM, Fundaci\'{o}n ``La Caixa'' and MICINN,  Spain;
SERI and SNSF,                               Switzerland;
T\"UB\.ITAK,                                 Turkey;
The Royal Society and UKRI/STFC,             United Kingdom;
DOE and NSF,                                 United States of America.

\bibliographystyle{apsrev4-2}
\bibliography{sectionbib}

\end{document}